\documentclass[journal]{IEEEtran} 
\usepackage{float}
\usepackage{stackengine}
\usepackage{setspace}
\usepackage{amssymb,amsfonts,latexsym}
\usepackage{graphicx}
\usepackage{subfigure}
\usepackage{psfrag}
\usepackage{algorithmic}
\usepackage{comment}
\usepackage{cite}
\usepackage{url}
\usepackage{times}
\usepackage{epsfig}
\usepackage{multirow}
\usepackage{longtable}
\usepackage{amsfonts}
\usepackage{amssymb}%
\usepackage{color}
\usepackage{mathrsfs}
\usepackage{bm}
\usepackage{booktabs}
\usepackage{threeparttable}
\usepackage[cmex10]{amsmath}
\usepackage{amsmath}
\usepackage[english]{babel}
\usepackage{mathbbold}
\usepackage[linesnumbered,ruled]{algorithm2e}
\makeatother
\usepackage[table]{xcolor}
\usepackage{makecell}
\usepackage{stfloats}

\newtheorem{proof}{\textbf{Proof}}

\newtheorem{remark}{Remark}
\newtheorem{lemma}{\textbf{Lemma}}

\ifCLASSINFOpdf
\else
\fi

\begin{document}
	\title{Beef up mmWave Dense Cellular Networks with \\D2D-Assisted Cooperative Edge Caching }
	
%
	\author{\small Wen Wu,~\IEEEmembership{\small Student~Member,~IEEE,}
	        Ning Zhang,~\IEEEmembership{\small Senior~Member,~IEEE,}
	        Nan~Cheng,~\IEEEmembership{\small Member,~IEEE,}\\
	        Yujie Tang,~\IEEEmembership{\small Member,~IEEE,}
	        Khalid Aldubaikhy,~\IEEEmembership{\small Student~Member,~IEEE,} and	        
	        Xuemin~(Sherman)~Shen,~\IEEEmembership{\small Fellow,~IEEE}
	\thanks{W. Wu, N. Cheng, Y. Tang, A. Khalid and X. Shen are with the Department of Electrical and Computer Engineering, University of Waterloo, Waterloo, ON N2L 3G1, Canada (e-mail:\{w77wu, n5cheng, y59tang, kaldubai, sshen\}@uwaterloo.ca). \emph{Corresponding author: Nan Cheng.} }
	\thanks{N. Zhang is with the Department of Computing Sciences, Texas A\&M University at Corpus Christi, TX, USA (email: ning.zhang@tamucc.edu).}}
	
\maketitle

\begin{abstract}
	Edge caching is emerging as the most promising solution to reduce the content retrieval delay and relieve the huge burden on the backhaul links in the ultra-dense networks by proactive caching popular contents in the small base station (SBS). However, constraint cache resource of individual SBSs significantly throttles the performance of edge caching. In this paper, we propose a \emph{device-to-device (D2D) assisted cooperative edge caching (DCEC)} policy for millimeter (mmWave) dense networks, which cooperatively utilizes the cache resource of users and SBSs in proximity. In the proposed DCEC policy, a content can be cached in either users' devices or SBSs according to the content popularity, and a user can retrieve the requested content from neighboring users via D2D links or the neighboring SBSs via cellular links to efficiently exploit the cache diversity. Unlike existing cooperative caching policies in the lower frequency bands that require complex interference management techniques to suppress interference, we take advantage of directional antenna in mmWave systems to ensure high transmission rate whereas mitigating interference footprint. Taking the practical directional antenna model and the network density into consideration, we derive closed-form expressions of the backhaul offloading performance and content retrieval delay based on the stochastic information of network topology. In addition, analytical results indicate that, with the increase of the network density, the content retrieval delay via D2D links increases significantly while that via cellular links increases slightly. Comprehensive simulations validate our theoretical analysis and demonstrate that the proposed policy can achieve higher performance in offloading the backhaul traffic and reducing the content retrieval delay compared with the state-of-the-art \emph{ most popular caching (MPC)} policy. 
	

	\vspace*{0mm}
	\begin{flushleft}
	\end{flushleft}	
\end{abstract}

\section{Introduction}

The proliferation of ever-increasing data-intensive wireless applications, such as virtual reality, augmented reality gaming, and high-definition video streaming, are expected to drastically strain the capacity of cellular networks in the foreseeable future \cite{xiao2017Jsac, cheng2018big}. To accommodate the surging demands of wireless data traffic, millimeter-wave (mmWave) communication, which is a de-facto candidate technology for on-going 5G networks, is envisaged to provide a pseudo-wire wireless connection service by exploiting a large swath of spectrum resource \cite{ rappaport2013it_will_work,cheng2017performance}. Leveraging high-gain directional antennas, current mmWave networks offer an extremely high data rate of nearly 7 Gbit/s, which is expected to increase to 40 Gbit/s in the forthcoming future. {As mmWave networks can be further densified due to the hostile propagation characteristics, deploying unconstrained wired backhaul links in dense networks becomes infeasible due to high costs, which results in backhaul congestion. Alleviating backhaul pressure is imperative for mmWave dense networks in the future.}

{Edge caching, which exploits the repetitive pattern of content requests in mobile applications, is a favorable solution to achieve this goal \cite{bastug2014living, zhong2017towards}. Specifically, popular contents can be cached in the small base station (SBS) during off-peak hours to serve users in proximity during peak hours, which has a potential to reduce up to 35\% backhaul traffic \cite{ye2017distributed, edgeCachingWhitePaper, SS-MAC}. In addition, edge caching provides a low-latency service since the content is retrieved from the edge instead of remote servers. However, due to the limited cache capacity of an individual SBS, the performance of edge caching can be constrained. To enlarge cached contents, a straightforward method is to leverage caching resources in a cooperative manner, i.e., cooperative caching. Cooperative caching can be divided into two categories: a) \emph{cooperative edge caching} where contents are cached in the SBS cluster which consists of multiple SBSs in proximity, and b) \emph{device-to-device (D2D) caching} where contents are cached in nearby users. In the cooperative edge caching, each SBS in the SBS cluster caches diverse contents to increase caching diversity and serves users, while in the D2D caching, each user and its neighboring user cache diverse contents and exchange cached contents via high-rate D2D communications.} 


Incorporating cooperative caching in mmWave dense networks can significantly relieve the backhaul burden and reduce the content retrieval delay. Besides, it also introduces an extra advantage. In lower frequency band systems, the performance of cooperative caching is throttled by multiuser interference which is caused by omni-directional transmission patterns, while directional antennas in mmWave systems naturally tackle the interference issue. However, cooperative caching in mmWave networks poses new challenges. There is no tractable analytical model for mmWave networks which incorporates the impacts of directional antenna, network density and content caching. Moreover, this intractability makes it difficult to attain closed-form expressions, which is arduous to provide valuable insights for the system design. 


In this paper, we propose a novel cooperative caching policy in mmWave cellular networks, which cooperatively utilizes cache resources of the user, its D2D peer and neighboring SBSs. Specifically, the contents are cached according to the content retrieval delay. The most popular contents are cached in the user and its D2D peer due to the low content retrieval delay, while less popular contents are cached in the SBS cluster. With such designed caching policy, we derive the backhaul offloading gain. In addition, we consider a practical mmWave directional antenna model, where the main lobe antenna gain varies and the side lobe antenna gain is non-zero, which entangles the interference analysis. Exploiting the stochastic information of the network topology, we theoretically analyze the average content retrieval delay of the proposed policy in mmWave systems. Analytical results reveal the impacts of the network density and practical directional antennas on caching performance, which are not well considered in the literature. Main contributions are summarized as follows:



\begin{itemize}
	\item A D2D-assisted cooperative edge caching (DCEC) policy which cooperatively exploits the cache resources of both users and SBSs, is proposed and analyzed in mmWave dense networks. 
	
	
	\item We derive closed-form expressions of the backhaul offloading gain and the content retrieval delay in mmWave dense networks based on stochastic information of network topology. 
	
	\item The impacts of the network density and practical directional antennas on caching performance are analyzed respectively. Analytical results show that content retrieval delay via D2D communications increases significantly with the network density, while that via cellular communications increases slightly. Besides, analytical results indicate the performance increases linearly with the directional antenna gain.
	\item Analytical results reveal the tradeoff relationship between transmission efficiency and caching diversity in mmWave dense networks, which investigates the optimal SBS cluster size.
	
\end{itemize}

The remainder of this paper is organized as follows. Section \ref{sec:literature review} reviews related works. Then, the system model is presented in Section \ref{sec:system model}. Next, we propose the DCEC policy and analyze its backhaul offloading gain. Section \ref{sec:network capacity analysis} analyzes the content retrieval delay performance of the proposed policy. Extensive simulations are presented in Section \ref{sec: simulation results}. Finally, concluding remarks are given in Section \ref{sec.Conclusion}.

\section{Literature Review}\label{sec:literature review}
{Endowed with the computing and storage functionalities, mobile edge computing (MEC) provides high quality of experience (QoE) for mobile users in proximity. A significant body of recent literature focuses on the computing functionality of MEC \cite{rodrigues2017hybrid, rodrigues2018cloudlets, zhou2017resource}. Rodrigues \emph{et al.} proposed a hybrid method through virtual machine migration and transmission power control, aiming at minimizing the service latency \cite{rodrigues2017hybrid}. An extended work focused on reconfiguring edge servers with an objective to improve the scalability \cite{rodrigues2018cloudlets}. Also, recently, the authors in \cite{zhou2017resource} jointly optimized the computing and caching resources to achieve the maximum utility in mobile edge networks. }

{Cooperative caching, which cooperatively utilizes storage functionality of MEC, is another approach to enhance QoE. Two methods are distinguished in the literature: D2D caching and cooperative edge caching. In microwave bands, both these two caching policies have been studied in an extensive body of work.} 
Firstly, by utilizing caching resources among users and high-rate D2D communications\cite{liu2016outage,song2018stable,zhou2017energy}, D2D caching can offload cellular traffic, increase cellular transmission rate and reduce the power consumption of SBS. 
A scaling law, where the throughput increases with the number of nodes in the network under an impractical condition that the D2D transmission range adjusts to the network density, is obtained in D2D caching networks \cite{ji2015throughput}. Wang \emph{et al.} investigated the performance of D2D caching in mobile scenarios where users frequently contact with neighboring users to exchange contents via D2D communications \cite{wang2017mobility}. In \cite{zhao2018caching}, three scheduling schemes for edge caching with D2D connections are proposed which can maximize
the throughput of D2D links with low complexity. 
Secondly, for the cooperative edge caching, caching resources in the SBS cluster are utilized to enlarge cached contents. Chen \emph{et al.} firstly presented a cooperative caching policy which cooperatively cached different fractions of less popular contents in different SBSs to increase content diversity, and revealed the tradeoff relationship between transmission diversity and content diversity \cite{chen2017cooperative}. To maximize the performance of the cooperative edge caching, content placement and cache size have been optimized. Zhang \emph{et al.} studied the delay-optimal problem via optimizing content placement, where a greedy algorithm is proposed to optimize content placement in the cooperative edge caching policy \cite{zhang2017cooperative}. Another work investigated the cache size optimization problem considering the budget of cache deployment in heterogeneous networks \cite{zhang2017cost}. Recent research in \cite{xu2018saving} applied in-memory storage and processing to enhance the energy efficiency of edge caching. {Also, recently, the authors in  \cite{zhao2018collaborative} developed a cooperative caching policy based on the content popularity distribution and user preference, which can improve the content hit ratio and reduce the transmission delay. Furthermore, observing the fact that popular contents are highly correlated to user locations, recent research \cite{yang2018content} proposed a location-aware caching policy through an online learning approach.}

Although interesting, these works solely focus on the cooperative caching policies operate in the microwave bands and do not consider the capability of mmWave communications. In addition, multiuser interference poses a challenge for cooperative caching at low frequency bands, especially in ultra-dense networks. Both D2D caching and cooperative edge caching require complex interference management technologies, such as power control and interference alignment, to reduce interference \cite{zhang2017cooperative, song2016packet}. However, this challenge can be easily addressed in mmWave systems as directional antennas significantly reduce multiuser interference. Studies on caching at mmWave frequency bands are quite limited. Semiari \emph{et al.} proposed a proactive caching policy to reduce handover failures in mobile mmWave networks \cite{semiari2018caching}. They focused on utilizing device caching and did not consider the cooperative edge caching. Ji \emph{et al.} first employed D2D caching in mmWave networks to enhance network performance \cite{ji2016wireless}. However, no analytical results in \cite{ji2016wireless} is provided to characterize the D2D caching performance. Then, in the very recent work \cite{giatsoglou2017d2d-aware}, Giatsoglou \emph{et al.} proposed and analyzed the D2D caching policy based on a stochastic geometry framework. However, this caching policy does not exploit cache resource of SBSs in proximity. In addition, the analytical results cannot characterize the impact of directional antennas in mmWave communications. Both solutions from \cite{ji2016wireless} and \cite{giatsoglou2017d2d-aware } solely applied D2D caching in mmWave networks to offload backhaul traffic while without taking the network density into consideration. Furthermore, cooperative edge caching policy and practical mmWave antenna features have not been investigated, which may greatly impact the network performance.

\begin{table}[t]
	\small
	\centering
	\caption{Variables and notations.}
	\label{Tab:Variables_and_notations}
	\begin{tabular}{l l l}
		\hline
		\hline
		\textbf{Notation} & \textbf{Description} \\
		\hline
		$\mathcal{F}$ & Requested file library\\
		$\Phi_{BS}$ & PPP of SBS\\
		$\lambda_{BS}$ & Density of SBS \\
		$\Phi_{UE}$ & PPP of users\\
		$\lambda_{UE}$ & Density of users\\
		$W$ & System bandwidth\\
		$\phi$ & Fraction of bandwidth allocation\\
		$\alpha$ &Path loss exponent\\
		$K$ & SBS cluster size\\
		$r$ & Physical distance \\
		$R$ & Transmission rate\\
		$\sigma^2$ & Background noise power\\
		$D$ & Average content retrieval delay\\
		$S$ & Signal power\\
		$I$ & Interference power\\
		$\xi$ & Content popularity skewness\\
		$G$ & Directional antenna gain\\
		$F$ & Backhaul offloading gain\\
		$h$ & Content hit ratio\\
		$\mathcal{B}_o$ & Associated SBS \\
		\hline
		\hline
	\end{tabular}
\end{table}	

\section{System Model}\label{sec:system model}
In this section, we present the network model, the content popularity model, the directional antenna model, the mmWave channel model and the transmission model, respectively. A summary of important notations is given in Table~\ref{Tab:Variables_and_notations}.

\subsection{Network Model}
\begin{figure}[t]
	\centering
	\renewcommand{\figurename}{Fig.}
	\includegraphics[width=0.4\textwidth]{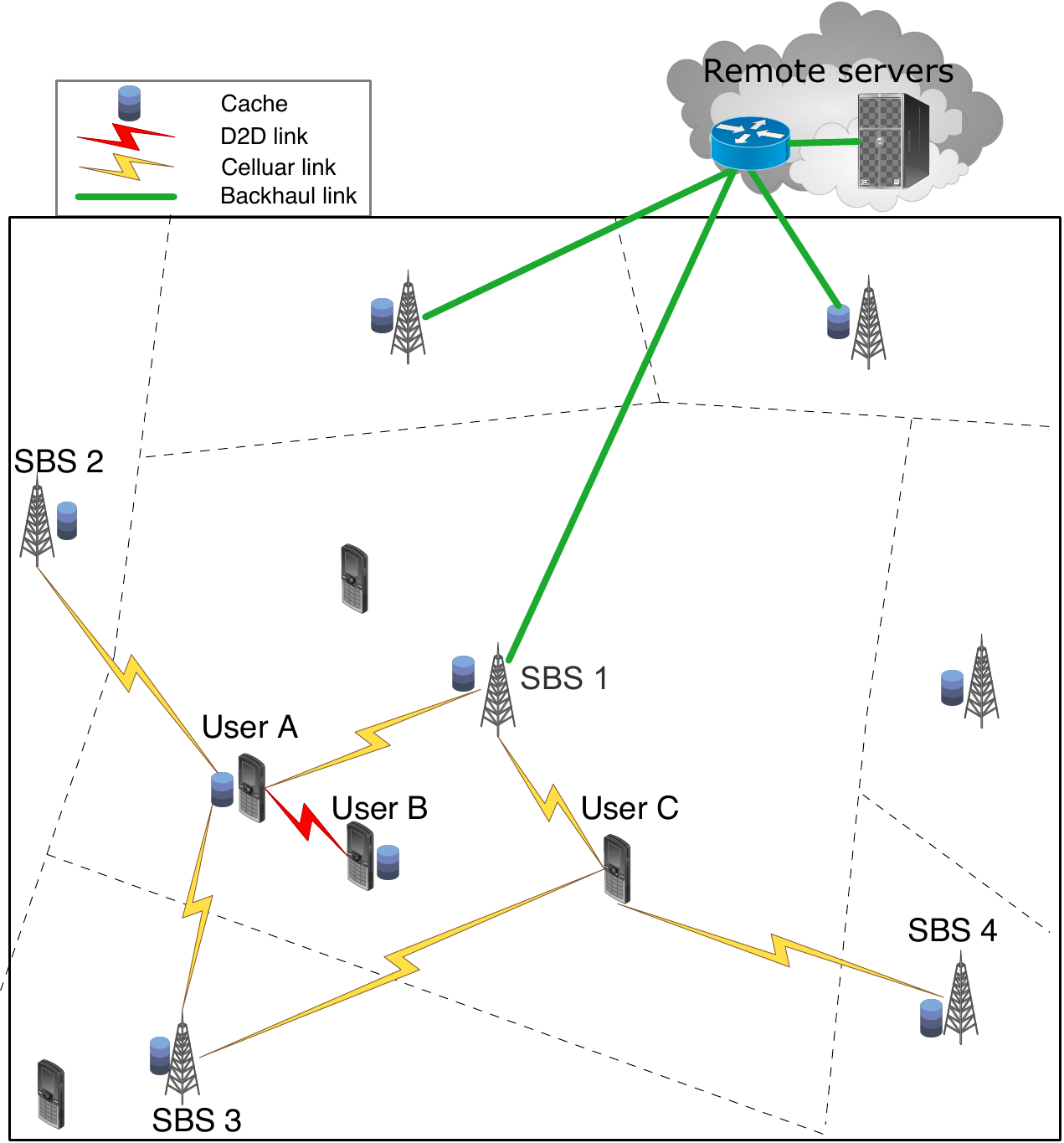}
	\caption{Cache-enabled edge network topology.}
	\label{Fig:network_topology}
\end{figure}

As shown in Fig. \ref{Fig:network_topology}, we consider a cache-enabled edge network where each entity can cache popular contents. SBSs and users follow homogeneous Poisson Point Processes (PPPs) $\Phi_{BS}$ and  $\Phi_{UE}$ on the plane, whose densities are given by $\lambda_{BS}$ and $\lambda_{UE}$, respectively \cite{giatsoglou2017d2d-aware,zhong2017heterogeneous}. {All the SBSs share the same spectrum and connect to remote servers with constrained backhaul links.} Each SBS adopts a time division multiple access (TDMA) mode to serve associated users. Both SBSs and users are equipped with steerable directional antennas. Beamforming training is perfectly performed between users and associated SBSs before the data transmission.

Considering the user-centric architecture \cite{zhang2017cooperative}, each user is allowed to be served by $K$ SBSs, which composes a SBS cluster, denoted by $\{ \text{SBS}_1,\text{SBS}_2,...,\text{SBS}_K \}$. For example, as shown in Fig. \ref{Fig:network_topology}, User $A$ is served by a SBS cluster with three SBSs, $\{\text{SBS}_1,\text{SBS}_2,\text{SBS}_3\}$. Users are divided into two categories: unpaired users and paired users. Unpaired users follow a homogeneous PPP $\Phi_{u}$ with a density of $\lambda_{u}$, which are only served by SBSs, such as User $C$ shown in Fig. \ref{Fig:network_topology}. A paired user is not only served by the SBS cluster but also its D2D peer. Paired users follow a homogeneous PPP $\Phi_{p}$ with a density of $\lambda_{p}$, and exchange cached contents via high-rate D2D communications. For example, User $A$ and User $B$ form a D2D pair and connect with each other via a D2D link. {For a paired user, its D2D peer uniformly distributes within a disk of radius $r_d^{max}$. Thus, the distance $r_d$ between the user and its D2D peer follows the following distribution \cite{giatsoglou2017d2d-aware}}
\begin{equation}\label{equ:D2D distance distribution}
f(r_d)=\frac{2r_d}{\left(r_d^{max}\right)^2}, 0<r_d<r_d^{max}.
\end{equation}

As D2D communications and cellular communications coexist in the system, we adopt an overlay scheme, i.e.,  D2D communications and cellular communications use different frequency bands to avoid interference. Assume that $W$ is the available bandwidth of the mmWave system, and $\phi W$ bandwidth is allocated to D2D communications. 

\subsection{Content Popularity Model}
Let $\mathcal{F}=\{f_1,f_2,...,f_i,... f_{|\mathcal{F}|}\}$ and $\mathcal{Q}=\{q_1,q_2,...,q_i,...,q_{|\mathcal{F}|}\}$ denote the sets of requested content and corresponding popularity distribution, respectively. $|\mathcal{F}|$ is the total number of contents. The Zipf distribution is used to characterize the popularity distribution \cite{zhang2017cooperative}, and the popularity of the $i$-th content is given by
\begin{equation}\label{equ:zipf distribution}
q_i=\frac{i^{-\xi}}{\sum_{j=1}^{|\mathcal{F}|} j^{-\xi}}, 1\leq i \leq |\mathcal{F}|
\end{equation}
where $\xi \geq 0$ denotes the content popularity skewness which varies based on content types. A larger popularity skewness value implies that the content requests are more concentrated.



\subsection{Directional Antenna Model}
Literature widely adopts the idealized ``flat-top" model, which has a constant antenna gain in the main lobe and zero elsewhere, to simplify the interference analysis \cite{singh2011interference}. However, in the practical directional antenna, the main-lobe antenna gain varies and the side-lobe antenna gain is non-zero, which brings difficulty in interference analysis and management. In this paper, we adopt a practical model with respect to a relative angle $\theta$ to its boresight to characterize the directional antenna gain, which is given by
\begin{equation}
G({\theta})=
\begin{cases}
G_m 10^{-c\left(\frac{2\theta}{\omega_m}\right)^2}  & |\theta| \leq \frac{\theta_m}{2}\\
G_s & \frac{\theta_m}{2} < |\theta| \leq \pi .
\end{cases}
\end{equation}
$G_m$ and $G_s$ denote the maximum antenna gain of the main lobe and the average antenna gain of the side lobe, respectively. $\omega_m$ and $\theta_m$ represent the beamwidth of the half-power and main lobe, respectively. $c$ is an experienced constant, which takes the value of $0.3$ \cite{xiao2017Jsac}. 

\subsection{mmWave Channel Model}
Regarding the channel model, the large-scale channel fading of mmWave links, in dB, is modeled as 
\begin{equation}\label{equ:path_loss_model}
PL(dB)=20\log_{10}\left(\frac{4\pi  d_{0} }{\lambda}\right)+10\alpha\log_{10}\left(\frac{r}{d_{0}}\right), r \geq d_0
\end{equation}
where $r$ and $\alpha$ denote the propagation distance and the path loss exponent, respectively. $\lambda$ is the wavelength and $d_0$ is the free space reference distance \cite{maccartney2017rural, wu2017Wiopt}. This model works well when the propagation distance is larger than the reference distance. For the sake of presentation, the average path loss can be rewritten as
\begin{equation}
\beta=C r^{-\alpha}
\end{equation} 
where $C=\frac{\lambda^2}{d_0^3 ({4\pi  })^2}$ is a constant.


For the small scale fading, the fast Raleigh fading is considered in this paper, i.e., $h\sim \exp(1)$ whose channel power gain is an exponential random variable with a unit mean. 

\subsection{Transmission Model}
The transmission rate of a mmWave link is given by
\begin{equation}\label{equ:transmission_rate}
R=\frac{W}{N_{cell}}\log_2\left(1+\frac{S}{I+\sigma^2}\right)
\end{equation}
where $N_{cell}$ is the cell load. The power of background thermal noise can be modeled as $\sigma^2={W}N_o $ where $N_o$ is the noise power spectral density. $S$ and $I$ represent the power of signal and interference, respectively.


In mmWave networks, each user is interfered by all the other SBSs excluding its associated SBS $\mathcal{B}_o$. Thus, considering the directional antenna model and the channel model, interference power is given by
\begin{equation}\label{equ:interference_simple}
\begin{split}
I &=\sum_{i\in \Phi_{BS} \backslash \mathcal{B}_o}I_i\\
&=\sum_{i\in \Phi_{BS} \backslash \mathcal{B}_o}P_B G(\theta_{t,i}) G(\theta_{r,i}) h_i C r_i^{-\alpha}
\end{split}
\end{equation}  
where $G(\theta_{t,i})$ and $G(\theta_{r,i}) $ represent transmit and receive directional antenna gains, receptively. $\theta_{t,i}$ and $\theta_{r,i}$ are the angle of departure (AOD) and the angle of arrival (AOA) of the interference link between the user and the $i$th interfered SBS, respectively. $r_i$ denotes the physical distance between the user and the $i$th interfered SBS. 
For tractability of the analysis, we assume that AOAs and AODs of interference links are uniformly distributed in $(0,2\pi]$ \cite{bai2015coverage}, which provides an average directional antenna gain for the interference signal. The average directional antenna gain is given by
\begin{equation}
\begin{split}
\bar{G}&=\int_{0}^{2\pi}G(\theta)f(\theta)d\theta\\
&=\int_{0}^{\frac{\theta_m}{2}}G_m 10^{-c\left(\frac{2\theta}{\omega_m}\right)^2} \frac{1}{\pi}d\theta+\int_{\frac{\theta}{2}}^{\pi}G_s \frac{1}{\pi}d\theta\\
&=\frac{\omega_m G_m}{\sqrt{2c\pi \ln {10}}}\text{erfc}\left(\frac{\theta_m \sqrt{c\ln10}}{\omega_m}\right)-\frac{G_s \theta_m}{2\pi}+G_s
\end{split}
\end{equation}
where $\text{erfc}(x)=\int_{0}^{x}e^{-t^2}dt$ represents the Gauss error function.
 Considering average directional antenna gains of interference links, the average interference power in \eqref{equ:interference_simple} can be rewritten as 
\begin{equation}\label{equ:interference_model}
\mathbb{E}[I]=\sum_{i\in \Phi_{BS} \backslash \mathcal{B}_o} P_B \bar{G}^2 C \mathbb{E}[h_i] \mathbb{E}[r_i^{-\alpha}].
\end{equation}
This interference model is used in the following analysis in this paper.
\section{D2D-Assisted Cooperative Edge Caching (DCEC) policy}\label{sec: proposed DACC scheme}
In this section, we first propose the DCEC policy to exploit caching diversity, and then analyze its backhaul offloading performance.
\subsection{Scheme Design}
In the DCEC policy, cache resource of the user, its D2D peer and the SBS cluster are utilized in a cooperative manner to store diverse contents to offload backhaul traffic. {Note that the user queries its D2D peer and the SBS cluster to identify where the requested content is cached.} For a requested content, if the content is cached in the user on-board storage, the user retrieves the content locally with negligible latency. Next, if the content is cached in its D2D peer, the user retrieves the content via D2D communications. Then, if the content is cached in an arbitrary SBS in the SBS cluster, the user associates to the SBS and then retrieves the content via cellular communications. Retrieving contents via cellular communications incurs higher delays compared with that via D2D communications, as D2D communications provide higher transmission rates than cellular communications due to shorter distances. Otherwise, if the content is miss cached, the user associates with the nearest SBS $B_o$ and retrieves the content from remote servers via the constraint backhaul link, which incurs a long delay. Thus, for a specific user, the content retrieval priority set which is sorted based on delay in an ascending order, is given by $\{$user $\leq$ its D2D peer $\leq$ SBS cluster  $\leq$ remoter servers$\}$. This content retrieval priority set is applied in the content placement of our proposed caching policy to minimize content retrieval delay, which is described in the following two steps. 

Firstly, the most popular contents are cached in both the user and its D2D peer due to the short content retrieval delay. 
Assume that each user has the same cache capacity, which is denoted by $C_u$. Note that users are divided into paired users and unpaired users. For paired users, the most popular $2C_u$ contents, i.e., $\{f_{1}, f_{2},...,f_{2C_u}\}$, are cached in the D2D pair which consists of the user and its D2D peer. For fairness, these $2C_u$ contents are equally distributed in the user and its D2D peer based on content popularity so that two users in the D2D pair have nearly the same content hit ratio. For example, User A and User B in the D2D pair have the same content hit ratio, i.e., $h_A=h_B$. 
Thus, the content hit ratio of a paired user is given by
\begin{equation}
h_p=\frac{1}{2}\sum_{i=1}^{2C_u} q_i.
\end{equation}   
For unpaired users, without the assistance of D2D peers, unpaired users can only cache the most popular $C_u$ contents, i.e., $\{f_{1}, f_{2},...,f_{C_u}\}$. Thus, the content hit ratio for an unpaired user is 
\begin{equation}
h_u=\sum_{i=1}^{C_u} q_i.
\end{equation}   

Secondly, less popular contents are cached in the SBS cluster due to the long content retrieval delay via cellular communications. Note that users are served by a SBS cluster with a size of  $K$. Assume that each SBS has the same cache capacity $C_s$. The SBS cluster caches the next $KC_s$ popular contents, i.e., $\{f_{2C_u+1}, f_{2C_u+2},...,f_{2C_u+KC_s}\}$. Similarly, for the fairness and cell load balance, these $KC_s$ contents are equally distributed in each SBS in the SBS cluster based on content popularity so that each SBS has the same content hit ratio, which is given by 
\begin{equation}
h_s=\frac{1}{K}\sum_{i=2C_u+1}^{2C_u+KC_s} q_i.
\end{equation}

Note that some fair popular contents are miss cached for the unpaired users. However, in the dense network scenario, the unpaired users only account for a small portion of all the users as users can associate with neighboring users with high probability. Specifically, in our simulations, we assume that 80\% users are paired users while only 20\% users are unpaired users. Hence, the impact of unpaired users on all the users is relatively small.


\subsection{Backhaul Offloading Analysis}
With the proposed caching policy, backhaul burden can be significantly relieved as users retrieve cached contents in edge networks instead of constrained backhaul links. In this paper, we define the \emph{backhaul offloading gain} as the ratio between data traffic that is not served by backhaul links and all the data traffic. In this subsection, the backhaul offloading gain of DCEC policy is analyzed.

For paired users, cache capacities of two users and the SBS cluster are cooperatively utilized to store the most popular $2C_u+KC_s$ contents, and hence the backhaul offloading gain is $2h_p+Kh_s$. For unpaired users, the backhaul offloading gain is $h_u+Kh_s$ without the assistance of the D2D peer. Thus, the average backhaul offloading gain of the DCEC policy is given by
\begin{equation}
\begin{split}
F&=h_u(1-\delta)+2h_p\delta+Kh_s
\end{split}
\end{equation}
where $\delta=\frac{\lambda_{p}}{\lambda_{p}+\lambda_u}$ denotes the fraction of paired users among all the users. It is obvious that the backhaul offloading gain increases with the cluster size $K$ due to the caching diversity gain. 

%
The corresponding miss caching probability, i.e., the probability that the requested content is not cached in edge networks, is given by
\begin{equation}
P_m=1-F.
\end{equation}
The miss cached contents can be retrieved from remote servers via the nearest cellular link and the constraint backhaul link. 


\section{Content Retrieval Delay Analysis}\label{sec:network capacity analysis}
In this section, the average \emph{content retrieval delay} of DCEC policy is analyzed. As users retrieve the requested content via different communication links, transmission performance of these communication links are analyzed respectively to obtain the average content retrieval delay. If the requested content is miss cached, users retrieve the content via two communication links. Firstly, the content is downloaded from remote servers to the nearest SBS with the average backhaul transmission rate $\mathbb{E}[R_B]$, and then transmitted to the user with the average nearest SBS transmission rate $\mathbb{E}[R_N]$. If the content is cached in the SBS cluster or its D2D peer, the user retrieves the content with the average SBS cluster transmission rate $\mathbb{E}[R_C]$ or the average D2D transmission rate $\mathbb{E}[R_D]$. Assume that the average content size is represented by $\nu$, the average content retrieval delay of the DCEC policy is given by
\begin{equation}
D=\frac{P_m\nu}{\mathbb{E}[R_B]}+\frac{P_m\nu}{\mathbb{E}[R_N]}+\frac{P_s\nu}{\mathbb{E}[R_C]}+\frac{P_d\nu}{\mathbb{E}[R_D]}
\end{equation}
where $P_s=Kh_s$ and $P_d=\delta h_p$ denote the probabilities that the requested content is cached in the SBS cluster and the D2D peer, respectively. In the following, lower bounds of these transmission rates are derived analytically respectively, and hence the upper bound of the average content retrieval delay is provided. 
\subsection{Backhaul Transmission Rate Analysis}
In this subsection, the average backhaul transmission rate is analyzed. According to the property of PPP, the traffic of each part in the network also follows PPP. Users served by constraint backhaul links are considered as a homogeneous PPP $\Phi_{B}$ in the plane with a density of $P_m \lambda_{UE}$. Assume each SBS has the same backhaul capacity $B$ and is served by its associated users with a TDMA mode. 

\begin{lemma}\label{lemma:backhaul capacity}
The average backhaul transmission rate of each user is given by
	\begin{equation}
	\mathbb{E}[R_B]=	\frac{B\lambda_{BS}}{ P_m \lambda_{UE}}\frac{\left(1+\frac{P_m \lambda_{UE}}{\kappa \lambda_{BS}}\right)^{\kappa+1}}{\left(1+\frac{P_m \lambda_{UE}}{\kappa \lambda_{BS}}\right)^{\kappa+1}-1}
	 \end{equation}
	 where $\kappa=3.5$ is a constant. 
\end{lemma}
\begin{proof}
As the backhaul resource is equally allocated to each user, the average backhaul rate of each user is $B/{\mathbb{E}[N_B]}$. $N_B$ is the backhaul load which is a random variable depending on the SBS cell area. The SBS cell area follows a Gamma distribution with a parameter $\kappa$. The probability distribution function (PDF) of the cell area $a$ is given by \cite{yu2013downlink}
\begin{equation}
f_{{}}(a)=a^{\kappa-1}e^{-\kappa\lambda_{BS}a}\frac{\left(\kappa\lambda_{BS}\right)^\kappa}{\Gamma\left(\kappa\right)}.
\end{equation} 	
	
Hence, the average backhaul load is
\begin{equation}\label{equ: average number of backhaul user}
\begin{split}
\mathbb{E}[N_B]&=\int_{a}^{\infty}\sum_{n=1}^{\infty}{n}\text{{Pr}}\{N_B=n|a\}f_{}(a)da\\
&\stackrel{{(a)}}{=}\int_{a}^{\infty}\sum_{n=1}^{\infty}{n}\frac{\left(P_m \lambda_{UE} a\right)^n}{n!}e^{-P_m \lambda_{UE} a}f_{{}}(a)da\\
&=\int_{a}^{\infty}{ P_m \lambda_{UE} a}\left(1-e^{- P_m \lambda_{UE} a}\right)\\
&a^{\kappa-1}e^{-\kappa\lambda_{BS}a}\frac{(\kappa\lambda_{BS})^\kappa}{\Gamma(\kappa)}da\\
&\stackrel{{(b)}}{=}\frac{P_m \lambda_{UE}}{\kappa\lambda_{BS} }\frac{\Gamma(\kappa+1)}{\Gamma(\kappa)}\left(1-\left(\frac{\kappa\lambda_{BS}}{\kappa\lambda_{BS}+P_m \lambda_{UE}}\right)^{\kappa+1}\right)\\
&=\frac{P_m \lambda_{UE}}{ \lambda_{BS}}\left(1-\frac{1}{\left(1+\frac{P_m \lambda_{UE}}{\kappa \lambda_{BS}}\right)^{\kappa+1}}\right).
\end{split}
\end{equation}
	
	$(a)$ follows from the fact that $N_B$ is a Poisson distribution random variable with a mean of $P_m \lambda_{UE} a$ \cite{haenggi2009stochastic}. $(b)$ is obtained via the definition of the gamma function $\Gamma(z)=\int_{0}^{\infty}x^{z-1}e^{-x}dx$. Thus, {Lemma~\ref{lemma:backhaul capacity}} is proved.
\end{proof}

\subsection{Nearest SBS Transmission Rate Analysis}
When the requested content is downloaded from remote servers via backhaul links, the user retrieves the content by associating to the nearest SBS, which provides the maximum cellular transmission rate. In this subsection, we aim to analyze the transmission rate of the nearest SBS. 

Since the overlay scheme is adopted in the system, $(1-\phi)W$ bandwidth is allocated to cellular communications and each SBS serves associated users with a TDMA mode. The associated users of each SBS consist of two categories: miss cached users whose requested content is miss cached and SBS cluster cached user whose requested content is cached in the SBS cluster. Thus, the associated users of each SBS are modeled as a PPP $\Phi_{C}$ with a density of $\lambda_C=(P_m+P_s) \lambda_{UE}$. The total number of SBS is given by $N_{BS}$. Similar to the result in \eqref{equ: average number of backhaul user}, the average cell load is given by
\begin{equation}
\mathbb{E}[N_C]=\frac{(P_m+P_s) \lambda_{UE}}{ \lambda_{BS}}\left(1-\frac{1}{\left(1+\frac{\left(P_m+P_s\right) \lambda_{UE}}{\kappa \lambda_{BS}}\right)^{\kappa+1}}\right).
\end{equation}


\begin{lemma}\label{Lemma: nearest SBS rate}
	When the user is associated to the nearest SBS $\mathcal{B}_o$, the average transmission rate is lower bounded by
	\begin{equation}
	\mathbb{E}[R_N] \geq \frac{(1-\phi)W}{\mathbb{E}[N_{C}]\ln2}\left(2\ln \frac{G_m}{\bar{G}}+\frac{(\alpha-2) \gamma}{2}-\ln J_1(\alpha)\right)
	\end{equation}
	where 
	\begin{eqnarray}
	\label{equ:J1(a)}
	J_1(\alpha) = \left\{\begin{matrix}
	\begin{split}
	&\frac{\Gamma(N_{BS}+1-\frac{\alpha}{2})}{(1-\frac{\alpha}{2})\Gamma(N_{BS})}-{\Gamma\left(1-\frac{\alpha}{2}\right)}, \mbox{                 } \alpha \neq 2\\ 
	&\ln(N_{BS}-1)+\gamma, \mbox{ } \alpha = 2
	\end{split}
	\end{matrix}\right.
	\end{eqnarray}
	and $\gamma$ is the Euler-Mascheroni constant whose value approximates to 0.577.
\end{lemma}

\begin{proof}
Assume directional antennas between the user and its associated nearest SBS are well-aligned, the received signal power is given by
\begin{equation}\label{equ: desired_signal_case1}
S=P_{B}G_m^2h_1Cr_1^{-\alpha}
\end{equation}
where $r_1$ is the distance between the user and the nearest SBS. The interference signal power includes the interference signal from all the other SBSs, and the interference signals are independent random variables \cite{Haenggi2010distance}. 

Using the transmission model in \eqref{equ:transmission_rate}, the average transmission rate is given by
\begin{equation}\label{lower bound of closest rate}
\begin{split}
\mathbb{E}[R_N]
&=\mathbb{E}\left[\frac{(1-\phi)W}{N_{C}}\log_2\left(1+\frac{S}{ I+\sigma^2}\right)\right]\\
&=\frac{(1-\phi)W}{\mathbb{E}[N_{C}]\ln2} \mathbb{E}\left[\ln\left(1+\frac{S}{\sum_{i\in \Phi_{BS} \backslash \mathcal{B}_o}I_i+\sigma^2}\right)\right]\\
&\geq \frac{(1-\phi)W}{\mathbb{E}[N_{C}]\ln2}\mathbb{E}\left[\ln \frac{S}{\sum_{i\in \Phi_{BS} \backslash \mathcal{B}_o}I_i}\right]\\
&\geq \frac{(1-\phi)W}{\mathbb{E}[N_{C}]\ln2}\left(\mathbb{E}[\ln S]-\ln\sum_{i\in \Phi_{BS} \backslash \mathcal{B}_o}\mathbb{E}\left[I_i\right]\right).
\end{split}
\end{equation}
The second step is because the received signal-to-interference-plus-noise-ratio (SINR) and the cellular load are independent random variables. The first inequality holds as the thermal noise can be ignored in high SNR scenarios. Practical applications are guaranteed with high SNR due to the reliable communication requirement. The last inequality holds from the Jensen inequality. In the following, $\mathbb{E}[\ln S]$ and $\sum_{i\in \Phi_{BS} \backslash \mathcal{B}_o}\mathbb{E}[I_i]$ are analyzed, respectively.


Firstly, substituting the definition of desired signal in \eqref{equ: desired_signal_case1}, $\mathbb{E}[\ln S]$ can be rewritten as
\begin{equation}\label{equ: desiredSignal final version}
\begin{split}
\mathbb{E}[\ln S]&=\mathbb{E}\left[\ln \left(P_{B}G_m^2h_1Cr_1^{-\alpha}\right)\right]\\
&=\ln \left(P_{B}G_m^2 C\right)+\mathbb{E}\left[\ln h_1\right]-\alpha \mathbb{E}\left[\ln r_1\right]\\
&=\ln \left(P_{B}G_m^2 C\right)-\gamma+\frac{\alpha}{2}\left( {\gamma+\ln\pi\lambda_{BS}}\right).
\end{split}
\end{equation}
 The equality holds because of the following two facts:
 $$\mathbb{E}[\ln h_1]=\int_{0}^{\infty}\ln x e^{-x}dx=-\gamma$$
 and
 	\begin{equation}
 	\begin{split}
 	\mathbb{E}[\ln r_1]&=\int_{0}^{\infty} \ln r_1f(r_1)dr_1\\
 	&\stackrel{{(a)}}{=}\int_{0}^{\infty} \ln(r_1) 2\pi \lambda_{BS} r_1 e^{-\pi \lambda_{BS}r_1^2}dr_1\\
 	&\stackrel{{(b)}}{=}\frac{1}{2}\left(\int_{0}^{\infty}e^{-y}\ln y dy- \int_{0}^{\infty}e^{-y}\ln(\pi \lambda_{BS}) dy\right)\\
 	&=-\frac{\gamma+\ln\pi\lambda_{BS}}{2}
  	\end{split}
 	\end{equation}
 	where $(a)$ is due to the fact that  $r_1$ obeys the following distribution $f(r_1)=2\pi \lambda_{BS} r_1 e^{-\pi \lambda_{BS}r_1^2}$ \cite{Haenggi2010distance}. $(b)$ follows by changing variable $y=\pi\lambda_{BS}r_1^2$.

Secondly, using the average interference model in \eqref{equ:interference_model}, $\sum_{i\in \Phi_{BS} \backslash \mathcal{B}_o}\mathbb{E}[I_i]$ is given as

\begin{equation}\label{equ:sum_interference}
\begin{split}
\ln\sum_{i\in \Phi_{BS} \backslash \mathcal{B}_o}\mathbb{E}[I_i]
&=\ln\sum_{i\in \Phi_{BS} \backslash \mathcal{B}_o}P_B \bar{G}^2 C\mathbb{E}[h_i]\mathbb{E}[r_i^{-\alpha}]\\
&=\ln\left(P_B \bar{G}^2C\right)+\ln \sum_{i\in \Phi_{BS} \backslash \mathcal{B}_o}\mathbb{E}[r_i^{-\alpha}].
\end{split}
\end{equation}
The last equality is due to the fact that $\mathbb{E}[h_i]=1$.

As the PDF of the distance between the user and $i$-th nearest SBS $r_i$ is given by \cite{haenggi2009stochastic}
$$
f(r,i)=\frac{2(\pi\lambda_{BS})^i}{(i-1)!}r^{2i-1}e^{-\pi \lambda_{BS}r^2}, i=2,3...
$$
the $-\alpha$th moments of $r_i$ can be calculated as follows:
\begin{equation}\label{equ:moments of r-a}
\begin{split}
\mathbb{E}[r_i^{-\alpha}]&=\int_{0}^{\infty}r^{-\alpha} f(r,i)dr\\
&=\int_{0}^{\infty}\frac{2(\pi\lambda_{BS})^i}{(i-1)!}r^{2i-1-\alpha}e^{-\pi \lambda_{BS}r^2}dr\\
&\stackrel{{}}{=}\frac{(\pi \lambda_{BS})^{\frac{\alpha}{2}}}{(i-1)!} \int_{0}^{\infty}y^{\frac{2i-\alpha}{2}-1}e^{-y}dy\\
&=(\pi \lambda_{BS})^{\frac{\alpha}{2}}\frac{\Gamma(i-\frac{\alpha}{2})}{\Gamma(i)},i>\frac{\alpha}{2}.
\end{split}
\end{equation}
Next, we aim to obtain the summation of the $-\alpha$th moments of $r_i$. 
When $\alpha$ attains different values, the summation of the $-\alpha$th moments of $r_i$ varies.
\begin{itemize}
	\item 
	When $\alpha \neq 2$, with results in \eqref{equ:moments of r-a}, the summation of the $-\alpha$th moments of $r_i$  is 
	\begin{equation}\label{equ: sum of moments of r_i 1}
	\begin{split}
	&\sum_{i\in \Phi_{BS} \backslash \mathcal{B}_o}\mathbb{E}[r_i^{-\alpha}]\\
	&=(\pi \lambda_{BS})^{\frac{\alpha}{2}} \sum_{i=2}^{N_{BS}}\frac{\Gamma(i-\frac{\alpha}{2})}{\Gamma(i)}\\
	&=(\pi \lambda_{BS})^{\frac{\alpha}{2}} \left(\frac{\Gamma(N_{BS}+1-\frac{\alpha}{2})}{(1-\frac{\alpha}{2})\Gamma(N_{BS})}-{\Gamma\left(1-\frac{\alpha}{2}\right)}\right).
	\end{split}
	\end{equation}
	The last equality follows from the following equality \cite{Haenggi2010distance}
\begin{equation}\label{equ:gamma sum }
\sum_{j=1}^{n}\frac{\Gamma(j-\beta)}{\Gamma(j)}=\frac{\Gamma(n+1-\beta)}{(1-\beta) \Gamma(n)}, \beta \neq 2.
\end{equation}
	
	\item 
	When $\alpha=2$, the summation is given by
	\begin{equation}\label{equ: sum of moments of r_i 2}
	\begin{split}
	\sum_{i\in \Phi_{BS} \backslash \mathcal{B}_o}\mathbb{E}[r_i^{-\alpha}]
	&=(\pi \lambda_{BS})^{\frac{\alpha}{2}}\sum_{i=2}^{N_{BS}}\frac{\Gamma(i-1)}{\Gamma(i)}\\
	&=(\pi \lambda_{BS})^{\frac{\alpha}{2}}\sum_{i=1}^{N_{BS}-1}\frac{1}{i}\\
	&\approx\left(\pi \lambda_{BS}\right)^{\frac{\alpha}{2}}\left(\ln(N_{BS}-1)+\gamma\right)
	\end{split}
	\end{equation}
\end{itemize}  
where the equality holds when $N_{BS}$ is large enough. Thus, this approximation is reasonable in dense networks.


Substituting \eqref{equ: sum of moments of r_i 1} and \eqref{equ: sum of moments of r_i 2} into \eqref{equ:sum_interference}, the logarithmic form of the summation of average interference power can be rewritten as
\begin{equation}\label{equ:sum interference final version}
\ln\sum_{i\in \Phi_{BS} \backslash \mathcal{B}_o}\mathbb{E}[I_i]= \ln P_B \bar{G}^2C+\frac{\alpha}{2}\ln \pi \lambda_{BS}+\ln J_1(\alpha)
\end{equation}
where the definition of $J_1(\alpha)$ is given in \eqref{equ:J1(a)}. Substituting \eqref{equ: desiredSignal final version} and \eqref{equ:sum interference final version} into \eqref{lower bound of closest rate}, {Lemma 2} is proved.
\end{proof}
\begin{remark}
	Lemma 1 characterizes the nearest SBS transmission performance in terms of system parameters, such as the network density, the directional antenna gain and the path loss exponent. Firstly, the average transmission rate increases linearly with the directional antennas gain, i.e., $G_m/\bar{G}$, which indicates that directional antennas enhance the throughput of mmWave systems. Secondly, the average transmission rate grows linearly with the path loss exponent $\alpha$ because hostile path loss at mmWave frequency bands severely mitigates interference and offers a spatial reuse gain. Thirdly, transmission performance slightly decreases with the network density as $\ln J_1(\alpha)$ slightly increases with the number of SBSs $N_{BS}$. The reason is that both communication distance and interference distance adjust to the network density.
\end{remark}


\subsection{SBS Cluster Transmission Rate Analysis}
Since each SBS in the SBS cluster has the same content hit ratio, the user has the same probability to associate to an arbitrary SBS in the SBS cluster. Thus, the average SBS cluster transmission rate should be averaged by transmission rates of all the candidate SBSs. 


\begin{lemma}\label{Lemma: cached SBS rate}
	The average SBS cluster transmission rate is given by
	\begin{equation}
	\begin{split}
	\mathbb{E}[R_C] &\geq\frac{(1-\phi)W}{\mathbb{E}[N_{C}]\ln2}\left(2\ln \frac{G_m}{\bar{G}}+\frac{(\alpha-2) \gamma}{2} \right.\\
	&\left. -\frac{\alpha}{2K}\sum_{k=1}^{K}\sum_{i=1}^{k-1}\frac{1}{i} -\frac{1}{K }\sum_{k=1}^{K}\ln J_2(\alpha,k)\right)
	\end{split}
	\end{equation}
	where
	\begin{eqnarray}
	\label{equ:J2(a)}
	J_2(\alpha,k) = \left\{\begin{matrix}
	\begin{split}
	& \frac{\Gamma(N_{BS}+1-\frac{\alpha}{2})}{(1-\frac{\alpha}{2})\Gamma(N_{BS})}-\frac{\Gamma(k-\frac{\alpha}{2})}{\Gamma(k)}, \mbox{                 } \alpha < 2\\ 
	& E_1(r_0)+ \ln(N_{BS}-1)+\gamma-J_4(k), \mbox{ } \alpha = 2\\
	&\Gamma\left(1-\frac{\alpha}{2} ,r_0\right)+\frac{\Gamma(N_{BS}+1-\frac{\alpha}{2})}{(1-\frac{\alpha}{2})\Gamma(N_{BS})}\\
	&-{\Gamma\left(1-\frac{\alpha}{2}\right)}-J_3(k) 
	\end{split}
	\end{matrix}\right.
	\end{eqnarray}
	
		\begin{eqnarray}
	\label{equ:J4(i)}
	J_3(k) = \left\{\begin{matrix}
	\begin{split}
	&\Gamma(1-\frac{\alpha}{2} ,r_0), \mbox{                 } k=1\\ 
	& \frac{\Gamma(k-\frac{\alpha}{2})}{\Gamma(k)}, \mbox{ } k\geq 2,
	\end{split}
	\end{matrix}\right.
	\end{eqnarray}
	\begin{eqnarray}
	\label{equ:J1(i)}
	J_4(k) = \left\{\begin{matrix}
	\begin{split}
	&E_1(r_0), \mbox{                 } k=1\\ 
	& \frac{1}{k-1}, \mbox{ } k\geq 2.
	\end{split}
	\end{matrix}\right.
	\end{eqnarray}
	Note that $r_0=\pi \lambda_{BS}d_{0}^2$. $\Gamma(z,a)=\int_{a}^{\infty}x^{z-1}e^{-x}dx$ and $E_1(x)=\int_{x}^{\infty}\frac{1}{t}e^{-t}dt$ denote the incomplete gamma function and the exponential integral function, respectively.
\end{lemma}

\begin{proof}
Let $\mathcal{B}_o=\{\mathcal{B}_o^1,  \mathcal{B}_o^2, ... ,\mathcal{B}_o^k, ...,  \mathcal{B}_o^K\} $ denote the set of candidate SBSs among the SBS cluster, which is sorted based on physical distances in an ascending order. The corresponding set of physical distances is $\{ r_1,r_2,...,r_k,...,r_K\}$.  

If the user is associated to the $k$th nearest SBS $\mathcal{B}_o^k$, the received desired signal power is given by
\begin{equation}
S_C^k=P_{B}G_m^2h_1Cr_k^{-\alpha}, 1 \leq k \leq K.
\end{equation}
The corresponding interference power consists of received signal power from all the SBSs excluding the $k$th nearest SBS, which is given by
\begin{equation}
I_C^k=\sum_{i\in \Phi_{BS} \backslash \mathcal{B}_o^k} P_B G(\theta_{t,i}) G(\theta_{r,i})  h_i C r_i^{-\alpha} , 1 \leq k \leq K.
\end{equation}

$R_C^k$ denotes the average transmission rate between the user and the $k$th nearest SBS. Hence, the average transmission rate among the SBS cluster can be represented by
\begin{equation}\label{equ:multiple SBS average rate}
\begin{split}
\mathbb{E}[R_C]&=\mathbb{E}\left[\frac{1}{K}\sum_{k=1}^{K}R_C^k\right]\\
&=\frac{(1-\phi)W}{K\mathbb{E}[N_{C}]\ln2}\sum_{k=1}^{K}\ln\left(1+\frac{S_C^k}{I_C^k+\sigma^2}\right)\\
&\geq \frac{(1-\phi)W}{K\mathbb{E}[N_{C}]\ln2}\sum_{k=1}^{K}\left(\mathbb{E}[\ln S_C^k]-\ln \sum_{i\in \Phi_{BS} \backslash \mathcal{B}_o^k} I_{i}^k\right)\\
&=\frac{(1-\phi)W}{\mathbb{E}[N_{C}]\ln2}\left(2\ln \left(\frac{G_m}{\bar{G}}\right)-\gamma-\frac{\alpha}{K} \sum_{k=1}^{K}\mathbb{E}[\ln r_k]\right. \\
&\left. -\frac{1}{K}\sum_{k=1}^{K}\ln\sum_{i\in \Phi_{BS} \backslash \mathcal{B}_o^k}\mathbb{E}[r_i^{-\alpha}]\right).
\end{split}
\end{equation} 
The inequality follows from the lower bound obtained in \eqref{lower bound of closest rate}. The last step follows from substitutions of \eqref{equ: desiredSignal final version} and \eqref{equ:sum_interference}. 
Then, according to \cite{zhang2017cooperative}, $\mathbb{E}[\ln r_k]$ is given by 
\begin{equation}\label{equ:SBS_cluster_log_r}
\begin{split}
\mathbb{E}[\ln r_k]
&=-\frac{1}{2}\left(\gamma+\ln(\pi\lambda_{BS})-\sum_{i=1}^{k-1}\frac{1}{i}\right).
\end{split}
\end{equation}

$\sum_{i\in \Phi_{BS} \backslash \mathcal{B}_o^k}\mathbb{E}[r_i^{-\alpha}]$ is derived in the following. When $\alpha$ attains different values, it can be represented in different forms.	
\begin{itemize}
	\item When $\alpha<2$, with results in \eqref{equ:moments of r-a}, we have
	\begin{equation}\label{equ:SBS_cluster_transmission_rate_a<2}
	\begin{split}
	\sum_{i\in \Phi_{BS} \backslash \mathcal{B}_o^k}\mathbb{E}[r_i^{-\alpha}]
	&=\sum_{i=1}^{N_{BS}}\mathbb{E}[r_i^{-\alpha}]-\mathbb{E}[r_k^{-\alpha}]\\
	&=(\pi\lambda_{BS})^{\frac{\alpha}{2}} \left(\frac{\Gamma(N_{BS}+1-\frac{\alpha}{2})}{(1-\frac{\alpha}{2})\Gamma(N_{BS})}\right.\\
	&\left.-\frac{\Gamma(k-\frac{\alpha}{2})}{\Gamma(k)}\right).
	\end{split}
	\end{equation}
	
	\item When $\alpha=2$,  $\mathbb{E}[r_1^{-\alpha}]=\int_{r_0}^{\infty}\frac{1}{y}e^{-y}dy=E_1(r_0)$
	according to the definition of the exponential integral function $E_1(x)$. While for $k\geq 2$, $\mathbb{E}[r_k^{-\alpha}]=\frac{1}{k-1}$ based on \eqref{equ: sum of moments of r_i 2}. Hence, $\mathbb{E}[r_k^{-\alpha}]$ can be described with a piecewise function $J_4(k)$ in \eqref{equ:J1(i)}. With the same method in \eqref{equ:SBS_cluster_transmission_rate_a>2}, we have 
	\begin{equation}\label{equ:SBS_cluster_transmission_rate_a=2}
	\begin{split}
	\sum_{i\in \Phi_{BS} \backslash \mathcal{B}_o^k}\mathbb{E}[r_i^{-\alpha}]
	&=\left(\pi\lambda_{BS}\right)\left(E_1(r_0)+ \ln(N_{BS}-1) \right.\\
	&\left.+\gamma-J_4(k)\right).
	\end{split}
	\end{equation}
	
	\item When $\alpha>2$, $\mathbb{E}[r_1^{-\alpha}]$ becomes unbounded because the condition $i>\frac{\alpha}{2}$ in \eqref{equ:moments of r-a} is not satisfied. The reason is that the path loss model in \eqref{equ:path_loss_model} breaks down when the distance is smaller than $d_{0}$. As a solution, we apply a guard radius $d_0$ among receivers to exclude interferers in the short distance. Hence, $-\alpha$th moments of $r_1$ becomes
\begin{equation}
\begin{split}
\mathbb{E}[r_1^{-\alpha}]&=\int_{d_0}^{\infty}r_1^{-\alpha} f(r_1)dr_1\\
&=(\pi \lambda_{BS})^{\frac{\alpha}{2}} \int_{\pi \lambda_{BS}d_{0}^2}^{\infty}y^{\frac{2-\alpha}{2}-1}e^{-y}dy\\
&=(\pi \lambda_{BS})^{\frac{\alpha}{2}}\Gamma(1-\frac{\alpha}{2} ,r_0).
\end{split}
\end{equation}
Thus, $\sum_{i\in \Phi_{BS} \backslash \mathcal{B}_o^k}\mathbb{E}[r_i^{-\alpha}]$ is given by
\begin{equation}\label{equ:SBS_cluster_transmission_rate_a>2}
\begin{split}
\sum_{i\in \Phi_{BS} \backslash \mathcal{B}_o^k}\mathbb{E}[r_i^{-\alpha}]
&=\mathbb{E}[r_1^{-\alpha}]+\sum_{i=2}^{N_{BS}}\mathbb{E}[r_i^{-\alpha}]-\mathbb{E}[r_k^{-\alpha}]\\
&=\left(\pi\lambda_{BS}\right)^{\frac{\alpha}{2}} \left(\Gamma(1-\frac{\alpha}{2} ,r_0)-{\Gamma(1-\frac{\alpha}{2})}\right.\\
&\left.+\frac{\Gamma(N_{BS}+1-\frac{\alpha}{2})}{(1-\frac{\alpha}{2})\Gamma(N_{BS})}-J_3(k)\right).
\end{split}
\end{equation}

\end{itemize}

Substituting \eqref{equ:SBS_cluster_log_r}, \eqref{equ:SBS_cluster_transmission_rate_a<2}, \eqref{equ:SBS_cluster_transmission_rate_a>2} and \eqref{equ:SBS_cluster_transmission_rate_a=2} into \eqref{equ:multiple SBS average rate}, {Lemma 3} is proved. 
\end{proof}

\begin{remark}
	{Lemma 3} gives the transmission performance of the SBS cluster with respect to mmWave system parameters and provides the following important observations. Similar to that in {Lemma 2}, the average SBS cluster transmission rate slightly decreases with network density because $J_2(\alpha,k)$ slightly increases with respect to the network density. {Lemma 3} indicates that the transmission rate decreases with the cluster size, which illuminates the tradeoff relationship between caching diversity and transmission efficiency. Enlarging cluster size increases cache capacity to cache more contents, while transmission performance degrades as users retrieve cached contents with a large distance. 
\end{remark}

\subsection{D2D Transmission Rate Analysis}
The D2D caching performance is analyzed in this subsection. D2D users follow a homogeneous PPP $\Phi_{D}$ and the corresponding density is $\lambda_{D}=P_d\lambda_{UE}$. Since mobile users provide a smaller directional antenna gain compared with SBSs due to limited space,  $G_u^m$ and $\bar{G}_u$ represent the maximal main lobe and the average antenna gains of users, respectively. Considering the overlay scheme, $\phi W$ system bandwidth is allocated to D2D communications. 
\begin{lemma}\label{Lemma: lemma4}
	The average D2D transmission rate is lower bounded by 
	\begin{equation}
	\begin{split}
	\mathbb{E}[R_D] &\geq \frac{{\phi W}}{\ln 2} 
	\left( 2\ln \frac{G_u^m}{\bar{G}_u} -\gamma-\alpha \left(\ln r_d^{max}-\frac{1}{2}\right) \right.\\
	&\left.  -\ln(\pi \lambda_{D}) -\ln J_5(\alpha)\right)
	\end{split}
	\end{equation}	
	where 
	\begin{eqnarray}
	\label{equ:J3(a)}
	J_5(\alpha) = \left\{\begin{matrix}
	\begin{split}
	&\frac{R^{2-\alpha}}{1-\frac{\alpha}{2}}, \mbox{                 } \alpha < 2\\ 
	&2  \ln \left( \frac{R}{d_0}\right), \mbox{ } \alpha = 2\\
	&\frac{R^{2-\alpha}-d_0^{2-\alpha}}{1-\frac{\alpha}{2}}, \mbox{                 } \alpha > 2
	\end{split}
	\end{matrix}\right.
	\end{eqnarray}
	 where $R=\sqrt{\frac{N_{D}}{\pi \lambda_{D}}}$ and $N_D$ is the number of D2D transmitters.

\end{lemma}
\begin{proof}
	The detailed proof is given in Appendix A.
\end{proof}
\begin{remark}
{Lemma 4} gives the transmission performance of D2D communication in terms of varying physical layer parameters. Most importantly, analytical results show that transmission performance decreases with the D2D user density $\lambda_{D}$ which depends on the network density. This is because that distance of interference links scales with the network density, while the D2D communication distance keeps unchanged. In addition, comparing {Lemma 2} and {Lemma 3} with {Lemma 4}, the average D2D transmission rate decreases significantly with the increase of the network density, while cellular transmission performance decreases slightly with the network density. 
Hence, the content retrieval delay via D2D communications increases significantly with the network density, which requires a coordinated scheduling scheme for D2D communications in dense networks. 
\end{remark}



\section{Simulation results}\label{sec: simulation results}
In this section, we compare the proposed DCEC policy with the state-of-the-art caching policy, and validate analytical results via extensive Monte-Carlo simulations. The simulation setup is given in Section \ref{subsec: simulation setup}. We evaluate the backhaul offloading performance in Section \ref{subsec:caching schemes comparison}, analytical results of transmission performance in Section \ref{subsec:analytical results evaluation} and content retrieval delay performance in Section \ref{subsec: delay evaluation}, respectively.

  \begin{table}[t]
	\small
	\centering
	\caption{Simulation parameters.}
	\label{Simulation parameters}
	\begin{tabular}{l l l l}
		\toprule
		\hline
		\textbf{Notation} &\textbf{Parameter} & \textbf{Value} \\
		\toprule
		$A$ & Simulation area &1 km$^2$\\
		$N_o$ &	Background noise density& $ -174$ dBm/Hz\\
		$W$   &Bandwidth & $2.16$ GHz\\
		$\phi$ & Fraction of D2D spectrum & 20\%\\
		$f$& Carrier frequency & $ 60$ GHz\\
		$\alpha$ & Path loss exponent & $\{1.4, 1.6, 2\}$\\
		$P_B$ & SBS transmit power & $30 $ dBm\\
		$P_U$ & User transmit power & $20 $ dBm\\
		$G_{s}^m$&SBS main lobe gain  & $18$ dB\\
		$G_{s}^s$&SBS side lobe gain  & $-2$ dB \\
		$G_{u}^m$&User main lobe gain  & $9$ dB\\
		$G_{u}^m$&User side lobe gain  & $-2$ dB \\
		$\omega_m$ &  Half-power beamwidth & $10^o$\\
		$d_0$ & Reference distance & $1$ m\\
		$r_d^{max}$ & Maximum D2D distance  & 10 m\\
		$\lambda_{BS}$ & Network density & \{80-400\} per km$^2$ \\
		$\lambda_{UE}$ & User density & \{800-4000\} per km$^2$ \\
		$\delta$ & Fraction of paired 		users & 80\% \\
		$F$ & Content library size & $2000$\\
		$\nu $ & Average content size & 100 Mbit\\
		$C_u$ & User cache capacity & 150\\
		$C_s$ & SBS cache capacity & 200\\
		\hline
		\hline
	\end{tabular}
\end{table}

\subsection{Simulation Setup}\label{subsec: simulation setup}

Important simulation parameters are listed in Table \ref{Simulation parameters}. We consider a simulation area of dimensions 1 km$^2$ (1000 m $\times$ 1000 m). Regarding the mmWave system, we consider the ratified IEEE 802.11 ad system which operates at the 60 GHz unlicensed band with a channel bandwidth of 2.16 GHz  \cite{wu2017performance}. 80\% bandwidth is allocated to cellular communications, while 20\% bandwidth to D2D communications. The parameters of directional antenna model are chosen based on typical values \cite{giatsoglou2017d2d-aware}. Here, we consider three different scenarios: the indoor conference room with line-of-sight (LOS) links, the living room with LOS links and none-line-of-sight (NLOS) links, and corresponding path loss exponents are set to 1.4, 1.6 and 2, respectively. Due to constraint on-board battery capacity and space in mobile devices, users adopt a lower transmit power and directional antenna gain compared to SBS. The constraint backhaul capacity is set to 3 Gbit/s unless otherwise specified. SBS density is chosen from 80 to 400 per km$^2$, with the average cell radius from 65 to 30 meters, corresponding to sparse rural and dense urban mmWave networks. The user density is set from 800 to 4000 per km$^2$, where 80\% users are paired users. We consider a requested content library with a total number of 2000 contents. The cache capacities of users and SBSs are set to 150 and 200 contents, respectively. To characterize video streaming applications, content popularity skewness is set to 0.56 unless otherwise specified \cite{zhang2017cost}.

In this section, we adopt the state-of-the-art most popular caching (MPC) policy as the benchmark. In the MPC policy, only the user and its associated SBS cache the most popular contents, i.e., the user and its associated SBS cache $\{f_1, f_2,...,f_{C_u}\}$ and $\{f_{C_u+1}, f_{C_u+2},...,f_{C_u+C_s}\}$, respectively. 
\subsection{Backhaul Offloading Performance}\label{subsec:caching schemes comparison}


\begin{figure*}[t]
	\centering
	\renewcommand{\figurename}{Fig.}
	\begin{subfigure}[Content popularity skewness.]{
			\label{Fig:offloadin_gain}
			\includegraphics[width=0.31\textwidth]{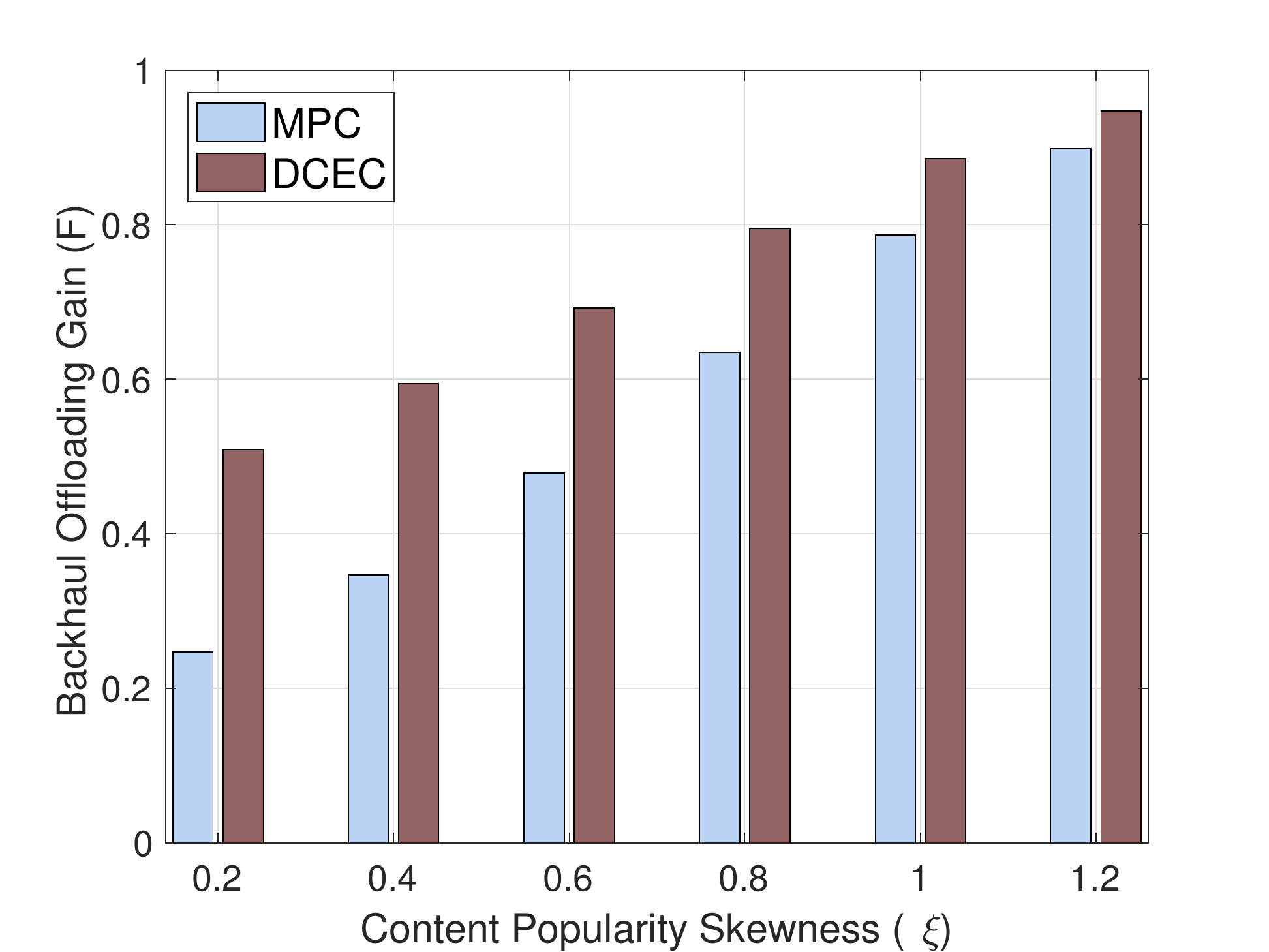}}
	\end{subfigure}%
	~
	\renewcommand{\figurename}{Fig.}
	\begin{subfigure}[SBS cache capacity]{
			\label{fig: differnt SBS cache capcity}
			\includegraphics[width=0.31\textwidth]{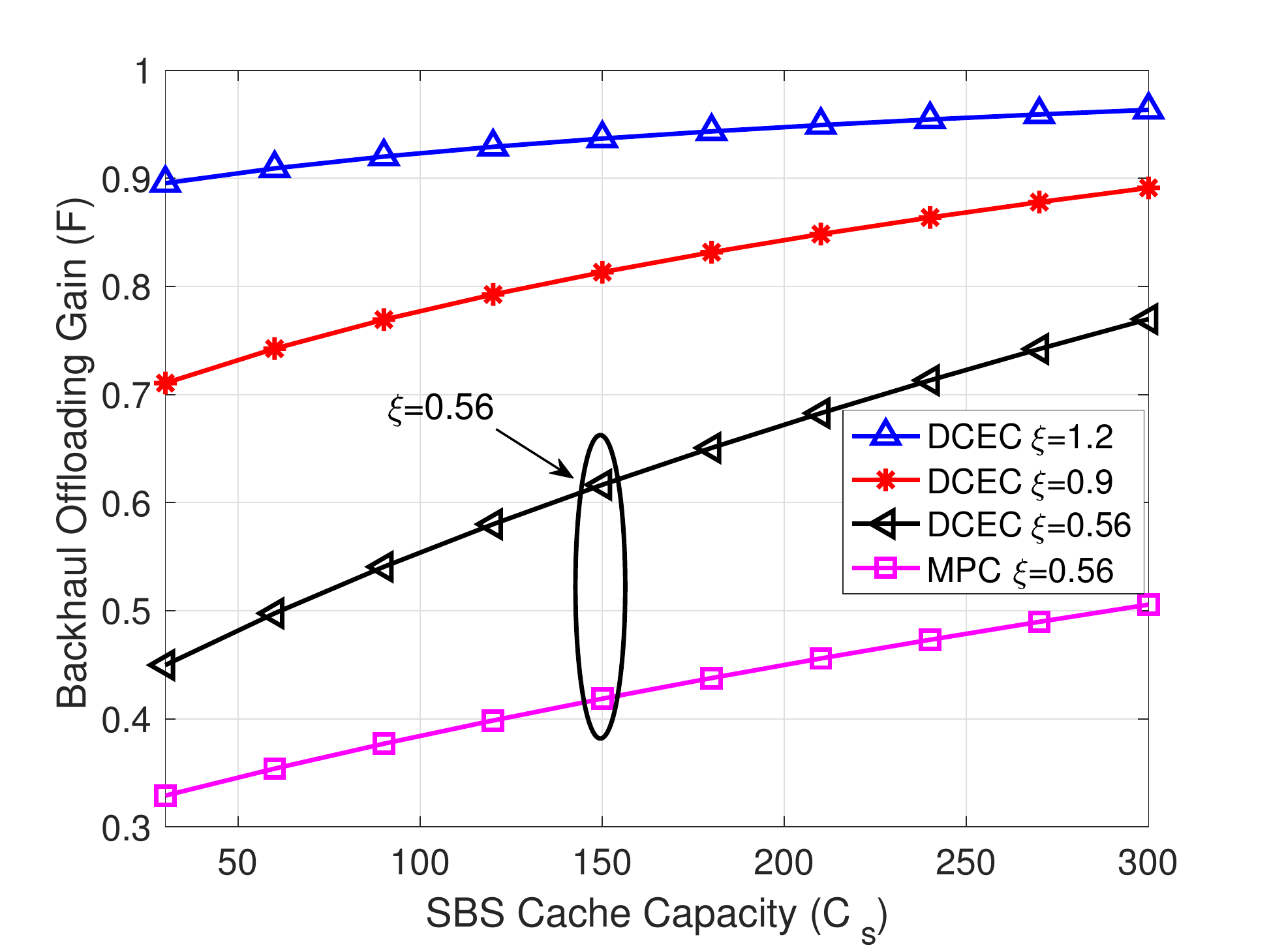}}
	\end{subfigure}%
	~
	\begin{subfigure}[SBS cluster size]{
			\label{fig:user_group_number_comparison}
			\includegraphics[width=0.31\textwidth]{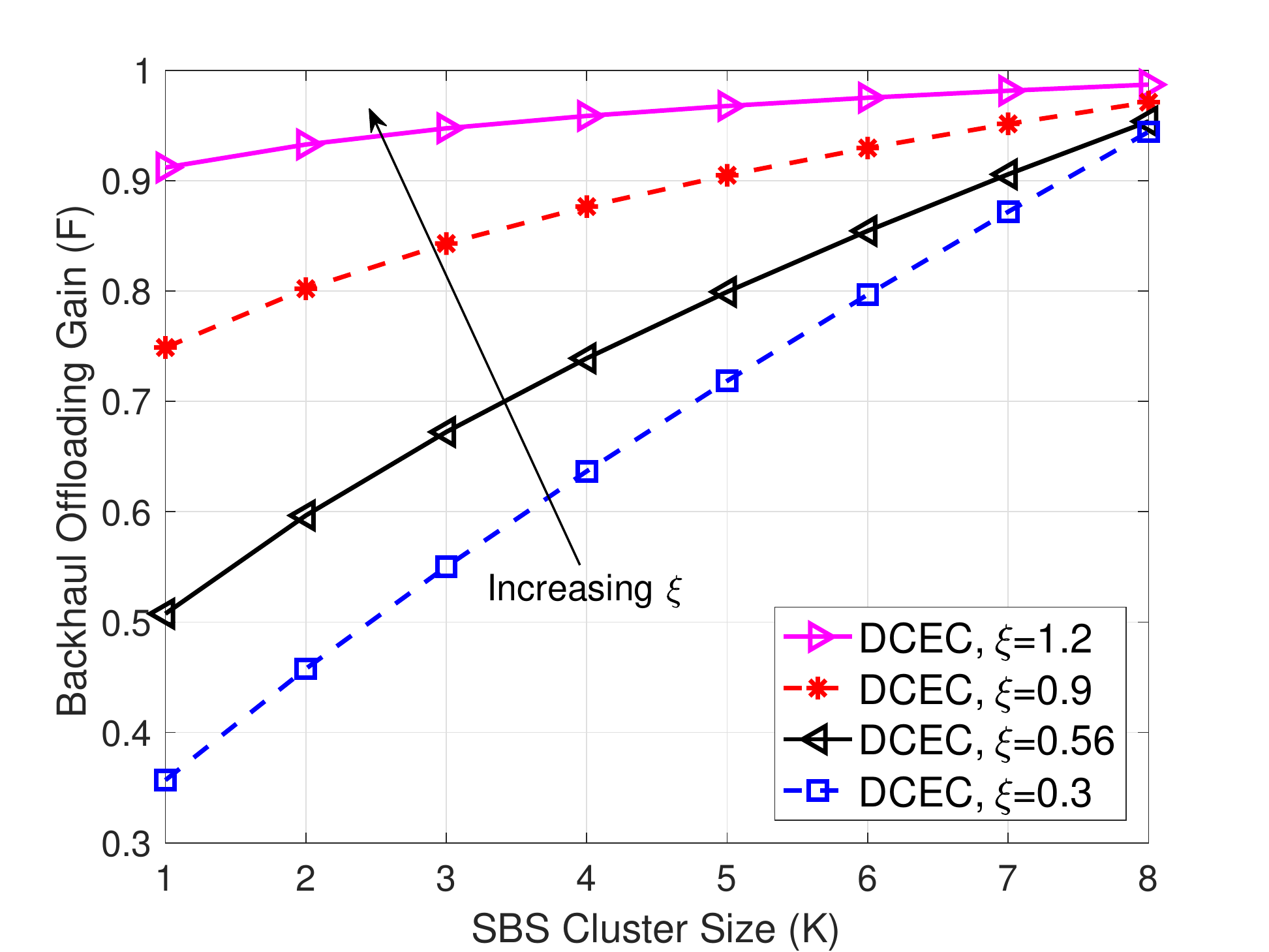}}
	\end{subfigure}
	\caption{Backhaul offloading performance with respect to different system parameters.}
	\label{Fig:offloading_comparison_with_different_parameters}
\end{figure*}


As shown in Fig. \ref{Fig:offloadin_gain}, we compare the backhaul offloading performance of the DCEC policy with the MPC policy in terms of popularity skewness. It is obvious that the proposed policy significantly outperforms the MPC policy due to exploiting cache resource efficiently. More specifically, the proposed policy offloads about 50\% backhaul traffic than MPC policy for $\xi=0.6$. Besides, the performance gap between these two policies narrows with the growth of the content popularity skewness. The reason is that the cache capacities of individual user and its associated SBS are enough to store highly concentrated contents.

Figure \ref{Fig:offloading_comparison_with_different_parameters} shows the backhaul offloading performance in terms of different system parameters. Fig. \ref {fig: differnt SBS cache capcity} shows the impact of SBS cache capacity on the backhaul offloading performance for $K=3$. We observe that more backhaul traffic is offloaded by local edge networks with the increase of the SBS cache capacity because more contents are cached. Specifically, caching resource growth provides more performance gain in the small popularity skewness region than that in large popularity skewness region. This is because that large cache capacity is favored for less concentrated content request applications. In addition, the performance gap between the DCEC policy and the MPC policy broadens with the increase of SBS cache capacity. We further demonstrate the backhaul offloading performance in terms of the SBS cluster size in Fig. \ref{fig:user_group_number_comparison}, where a significant backhaul offloading gain is achieved with the increase of SBS cluster size for low popularity skewness values. Particularly, the DCEC policy with eight SBSs offloads 70\% backhaul traffic more than that with two SBSs for $\xi=0.56$. But for large values of $\xi$, the performance enhancement becomes limited. Therefore, the DCEC policy favors less concentrated content request applications.

%
%

\subsection{Transmission Performance}\label{subsec:analytical results evaluation}

In this subsection, analytical results of transmission performance are validated via extensive simulations. The simulation results are averaged over 10000 samples with different network topologies and channel fading. 

%
%
%

\begin{figure*}[t]
	\centering
	\renewcommand{\figurename}{Fig.}
	\begin{subfigure}[The nearest SBS transmission rate ]{
			\label{Fig:closest_BS_average_rate}
			\includegraphics[width=0.4\textwidth]{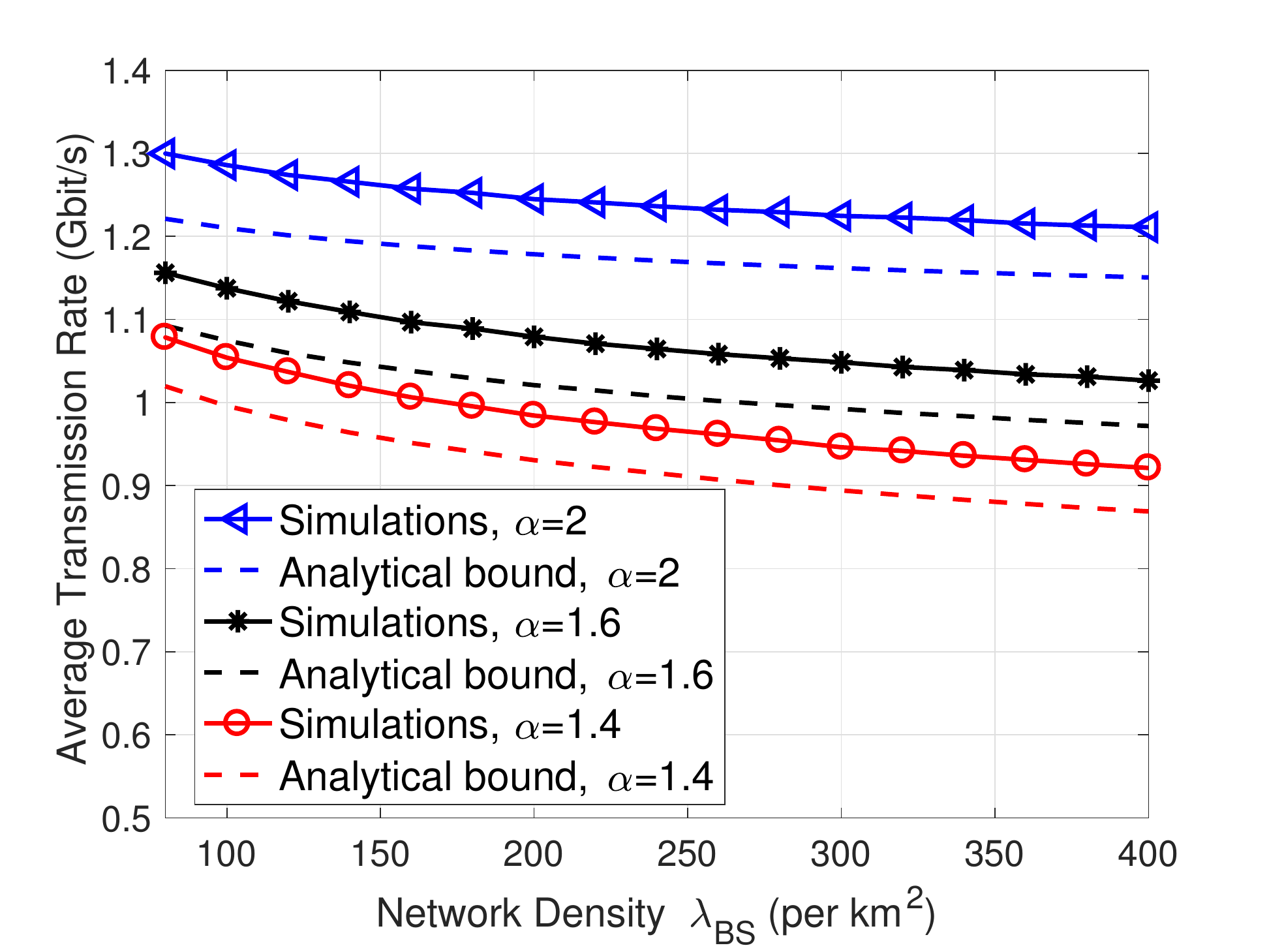}}
	\end{subfigure}%
	~
	\begin{subfigure}[SBS cluster transmission rate]{
			\label{Fig:multiple_BS_average_rate}
			\includegraphics[width=0.4\textwidth]{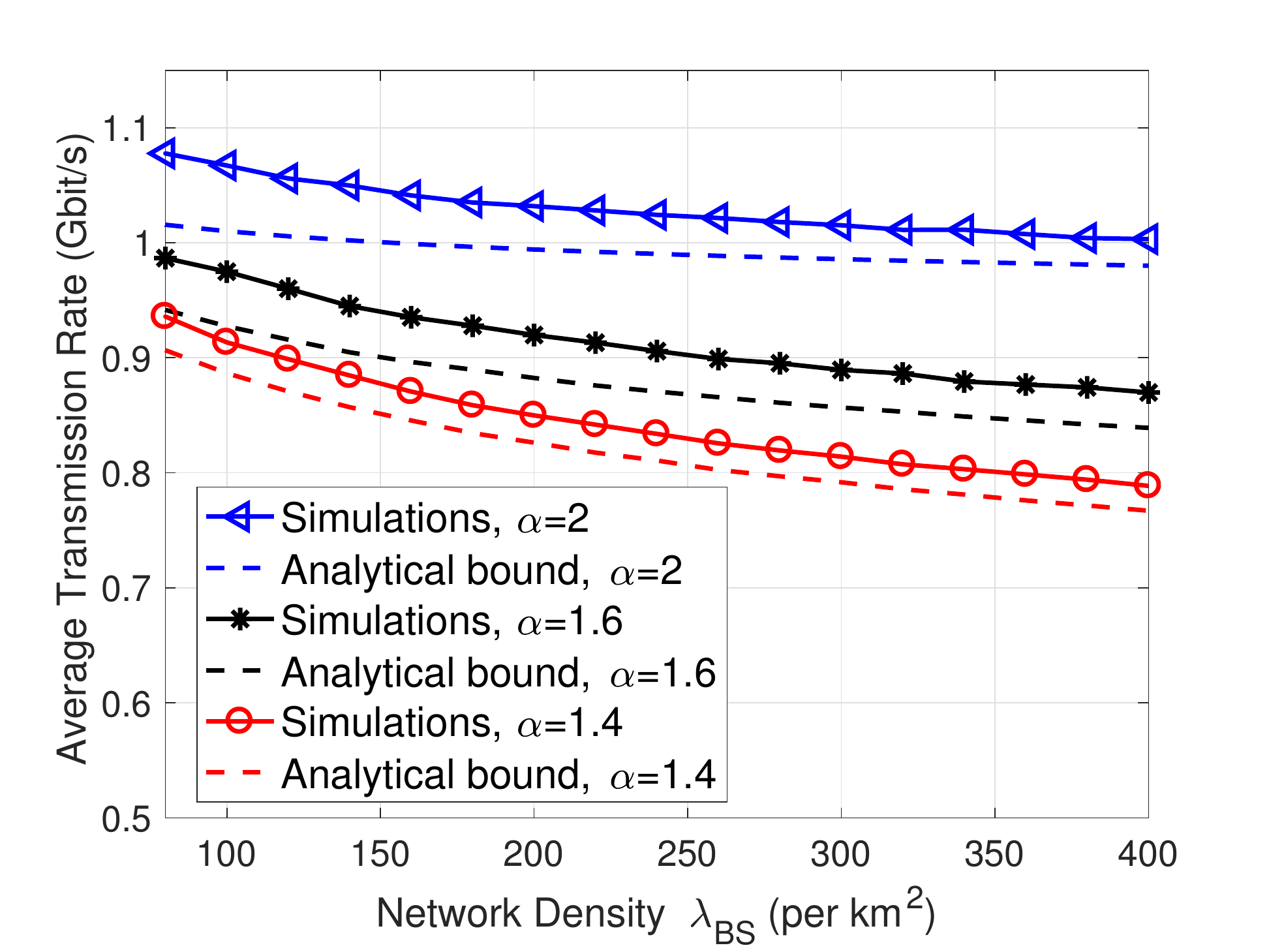}}
	\end{subfigure}
	~
	\begin{subfigure}[SBS cluster transmission rate with respect to the cluster size]{
			\label{Fig:multilpe_SBS_rate_versus_K}
			\includegraphics[width=0.4\textwidth]{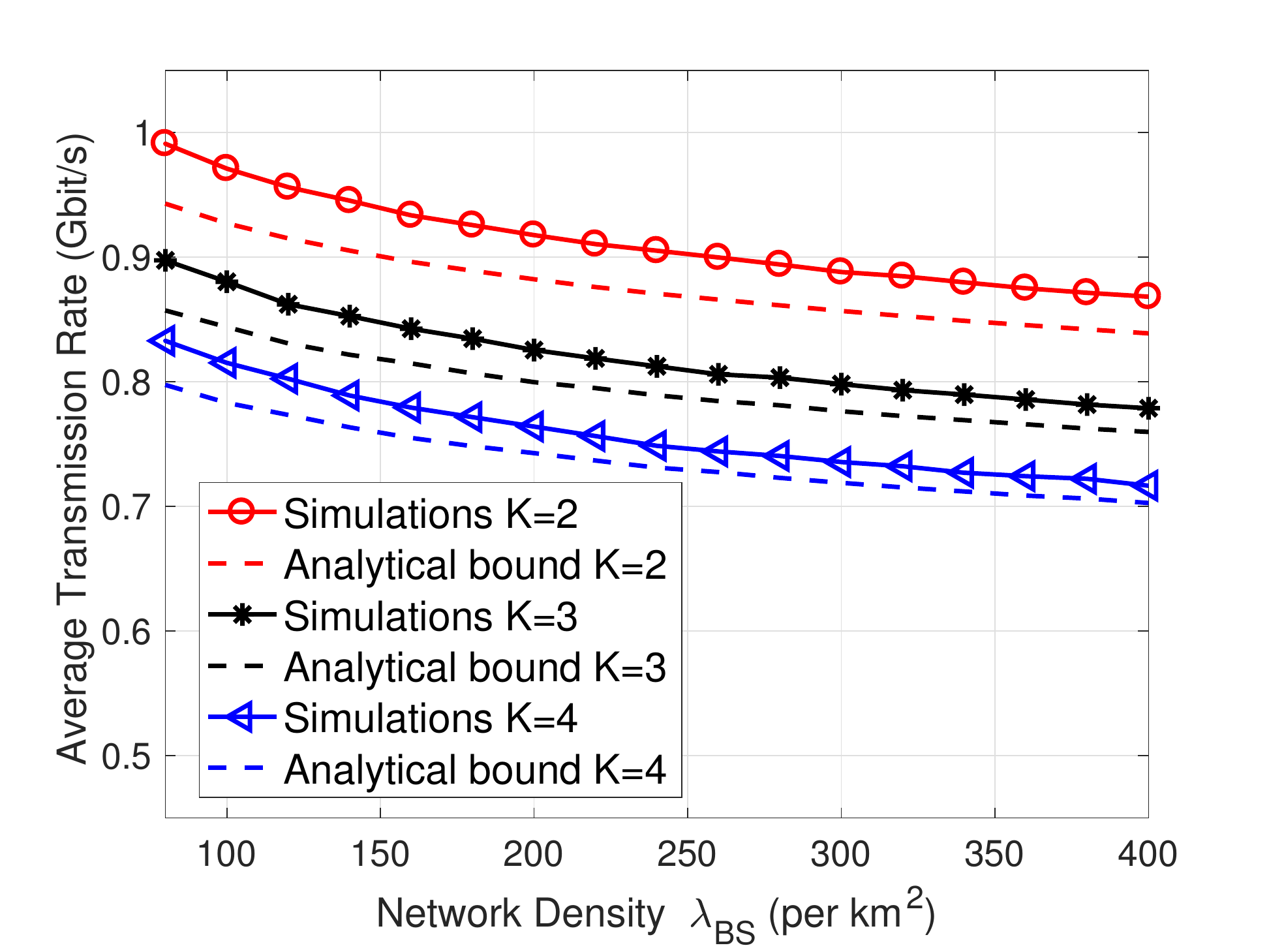}}
	\end{subfigure}
	~
	\begin{subfigure}[D2D transmission rate ]{
			\label{Fig:D2D peer average rate}
			\includegraphics[width=0.4\textwidth]{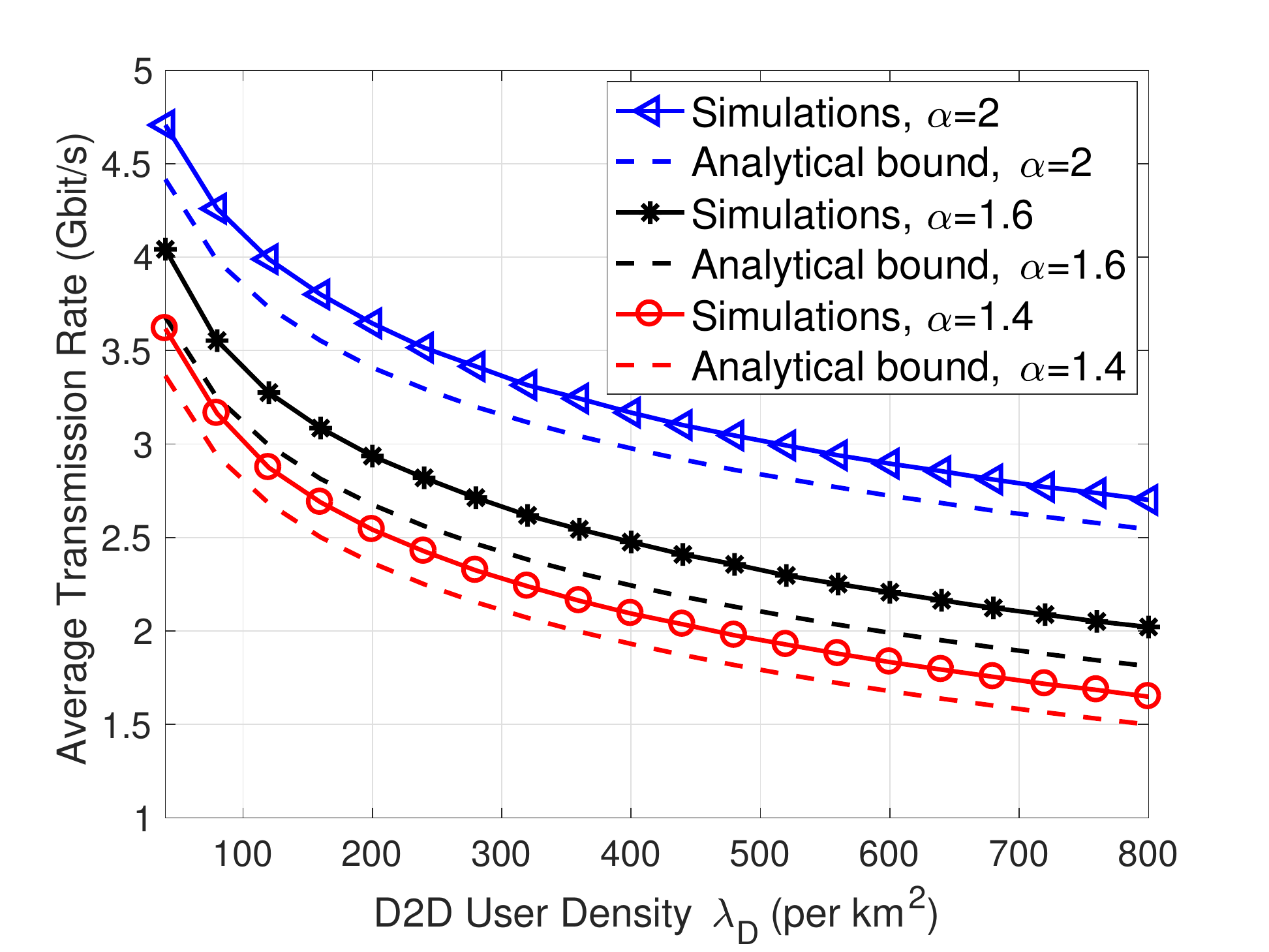}}
	\end{subfigure}
	\caption{Transmission performance with respect to different system parameters.}
	\label{Fig:transmission_performance}
\end{figure*}

Figure \ref{Fig:closest_BS_average_rate} shows the nearest SBS transmission rate in terms of the network density for $\alpha=1.4$, $1.6$, $2$. The analytical lower bounds in {Lemma 2} are quite close to simulation results under different channel conditions, which validates our analytical results. More importantly, the average rate slightly decreases with the network density. For $\alpha=2$, the transmission rate only decreases by 8\% as the network density increases from 80 to 400 per km$^2$. Even for $\alpha =1.4$, the transmission rate only degrades by 15\%. In addition, we observe that the average transmission rate increases with respect to $\alpha$ due to the fact that interference is alleviated by severe propagation loss in mmWave channels.

As shown in Fig. \ref{Fig:multiple_BS_average_rate}, we investigate the average SBS cluster transmission rate for $K=2$ and $\alpha=1.4$, $1.6$, $2$. Narrow gaps are observed between simulation results and analytical bounds with respect to the network density, which indicates that our analytical results in {Lemma 3} are quite accurate. Additionally, we observe that the transmission performance slightly decreases with the network density, which is similar as that of the nearest SBS. Particularly, for $\alpha =2$, the average transmission rate decreases by only 10\% from the sparse network for $\lambda_{BS}=80$ to the dense network for $\lambda_{BS}=400$. 

Figure \ref{Fig:multilpe_SBS_rate_versus_K} shows the average transmission rates in terms of the network density with different cluster sizes for $\alpha=1.6$. Analytical results are highly consistent with simulation results, which further corroborates the accuracy of {Lemma \ref{Lemma: cached SBS rate}}. More importantly, we observe that the average SBS cluster transmission rate decreases with the increase of the cluster size. Specifically, the SBS cluster with two SBSs provides a data rate is around 1.08 Gbit/s for $\lambda_{BS}=80$, while the SBS cluster with four SBSs only provides a data rate of 0.83 Gbit/s for the same network density, which decreases by nearly 23\%. This is because users retrieve cached contents from remote SBSs with long distances. A large SBS cluster caches more contents while reduces the average transmission rate, which reveals the tradeoff relationship between caching diversity and transmission efficiency.

In Fig. \ref{Fig:D2D peer average rate}, we investigate the average D2D transmission rate with respect to the D2D user density for $\alpha=1.4$, $1.6$, $2$. Simulation results match the analytical bounds in {Lemma \ref{Lemma: lemma4}} under different channel conditions. Firstly, it can be seen that D2D communications achieve a higher transmission rate compared with cellular communications due to short distances, which means a low content retrieval delay via D2D communications. Then, we observe a sharp performance degradation with the growth of the D2D user density, i.e., the transmission rate decreases from 4 Gbit/s to only 2 Gbit/s when the D2D user density increases from 40 to 800 per km$^2$ for $\alpha=1.6$. For this reason is that the desired signal power keeps unchanged as D2D communication distance is independent of D2D user density, while the interference increases drastically due to decreasing interference link distances with the D2D user density. Hence, a coordinated scheduling scheme is required to enhance D2D transmission performance in dense networks.

\subsection{Content Retrieval Delay}\label{subsec: delay evaluation}
We investigate the content retrieval delay performance with different caching policies with respect to the content popularity, the network density, the backhaul capacity and the cluster size, respectively. 
\begin{figure*}[t]
	\centering
	\renewcommand{\figurename}{Fig.}
	\begin{subfigure}[Content popularity]{
			\label{Fig:Delay_versus_xi.eps}
			\includegraphics[width=0.31\textwidth]{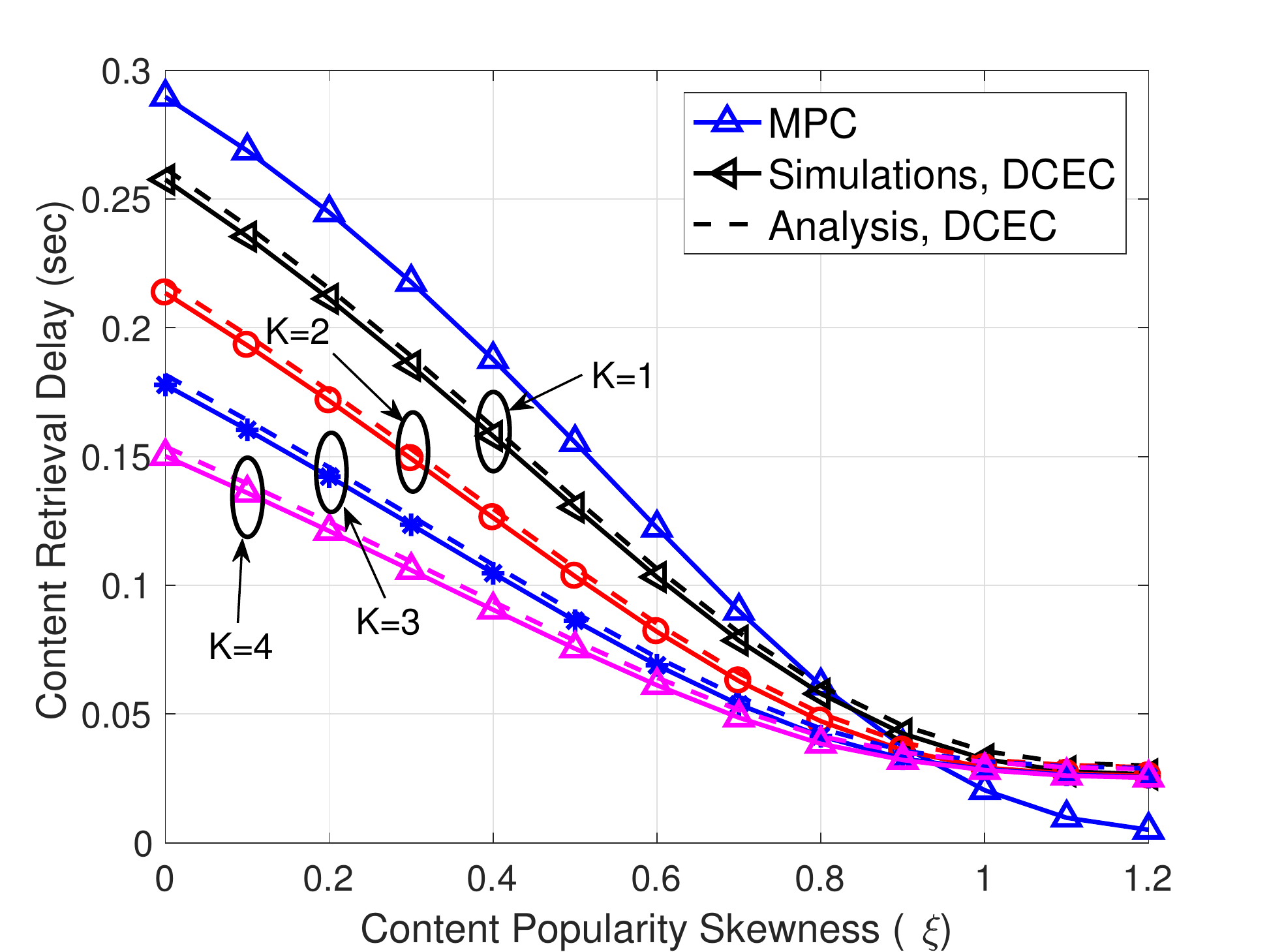}}
	\end{subfigure}%
	~
	\begin{subfigure}[Network density]{
			\label{Fig:delay_versus_user_density}
			\includegraphics[width=0.31\textwidth]{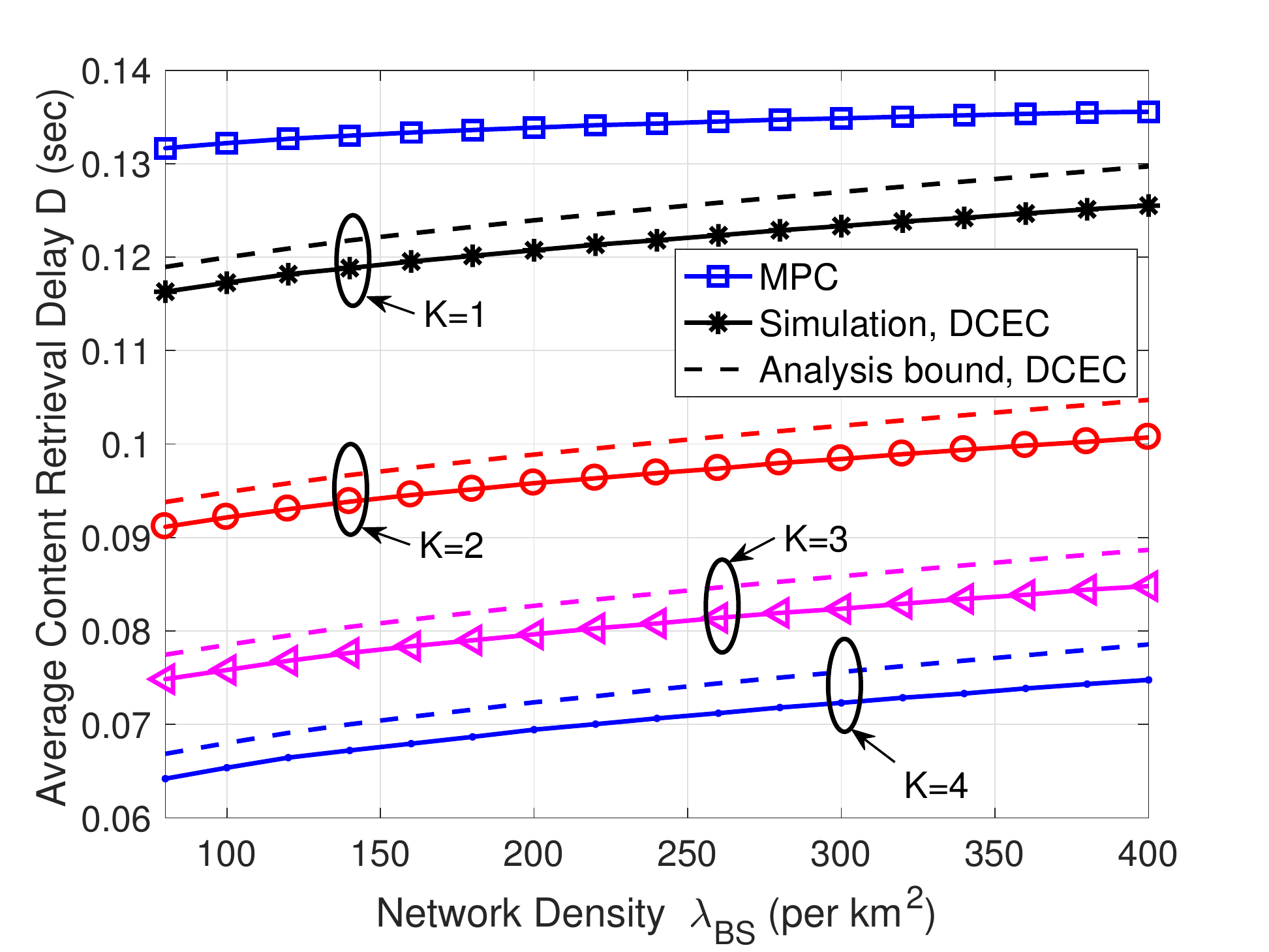}}
	\end{subfigure}
	~
	\begin{subfigure}[Backhaul capacity]{
			\label{Fig:delay_versus_backhaul_capacity}
			\includegraphics[width=0.31\textwidth]{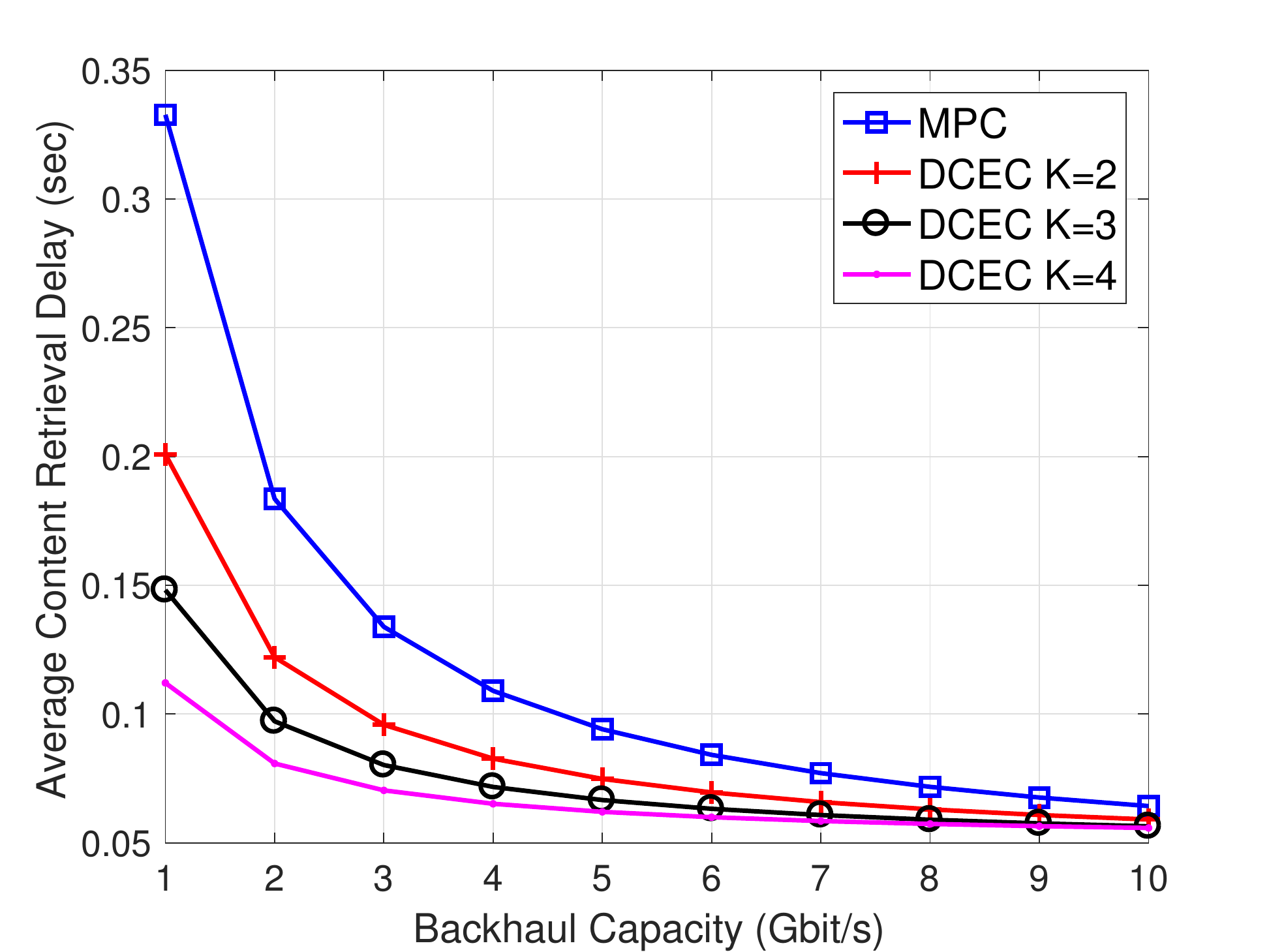}}
	\end{subfigure}
	\caption{Content retrieval delay with respect to different system parameters.}
\label{Fig:average_delay_performance}
\end{figure*}

 \begin{figure}[h]
	\centering
	\renewcommand{\figurename}{Fig.}
	\includegraphics[width=0.4\textwidth]{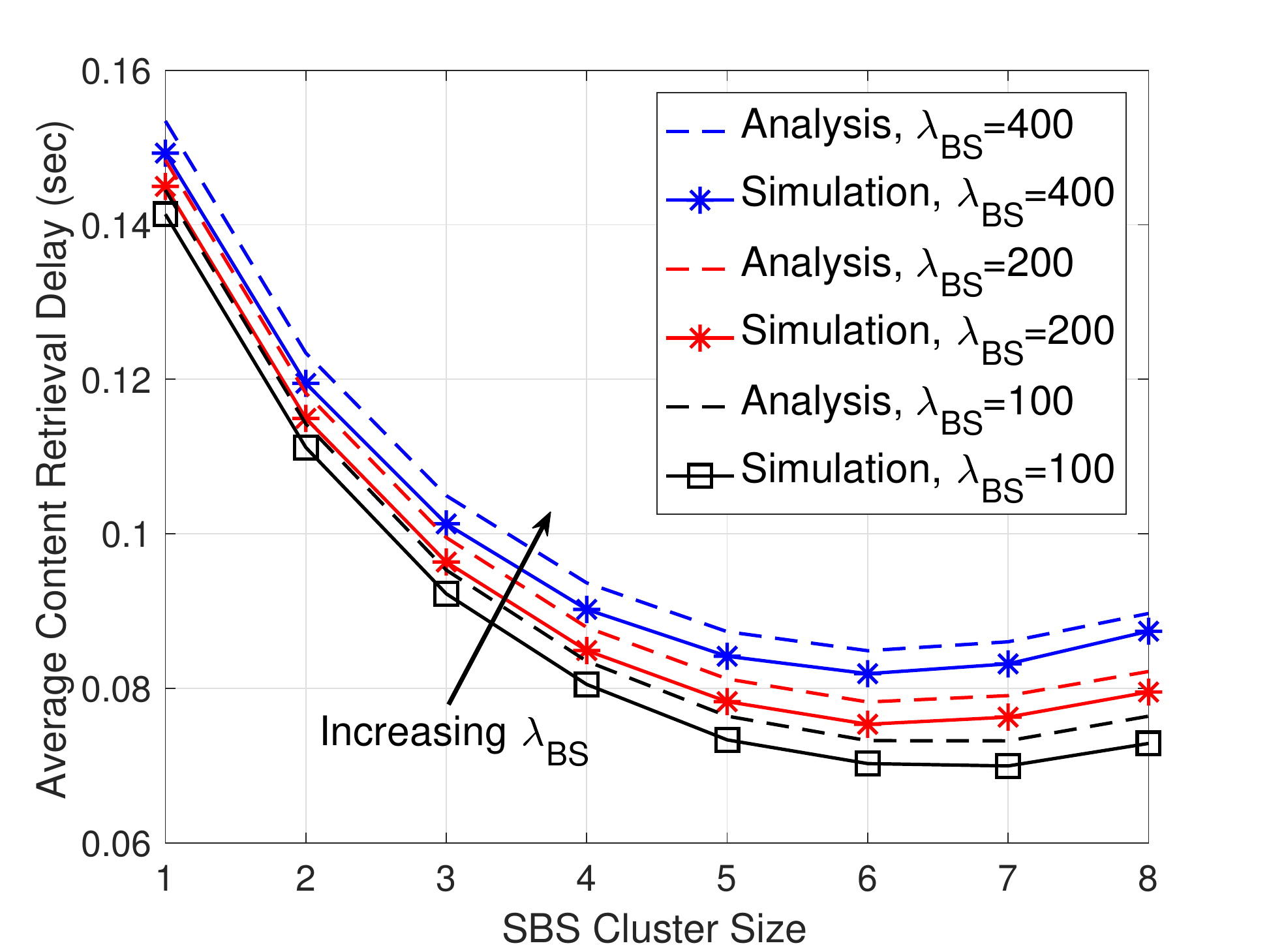}
	\caption{The impact of SBS cluster size on the content retrieval delay.}
	\label{Fig:tradeoff}
\end{figure}

 \begin{figure}[h]
	\centering
	\renewcommand{\figurename}{Fig.}
	\includegraphics[width=0.4\textwidth]{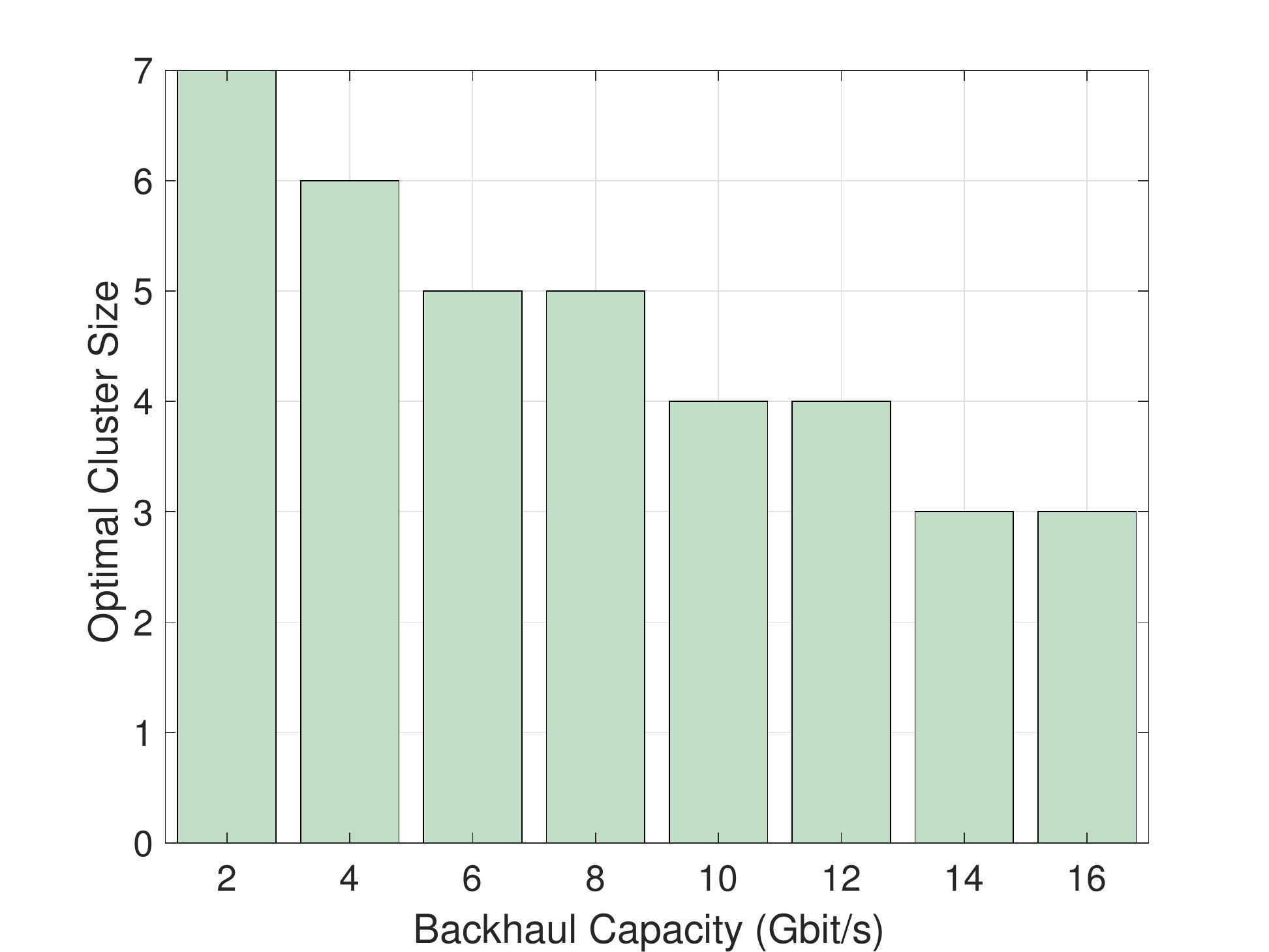}
	\caption{Optimal SBS cluster size with respect to backhaul capacity.}
	\label{Fig:delay_versus_cluster_size}
\end{figure}



Figure \ref{Fig:Delay_versus_xi.eps} shows the impact of the content popularity skewness on the average content retrieval delay. The proposed policy outperforms the MPC policy in the low popularity skewness region because low concentrated content request applications favor large cache capacity. For example, the DCEC policy with four SBSs reduces 48\% content retrieval delay compared with the MPC policy for $\xi=0.6$. However, the performance gain provided by the DCEC policy vanishes with the increase of the content popularity skewness, which further validates the fact that DCEC policy favors less concentrated content request applications. 
 
The content retrieval delay performance in terms of the network density is studied for $\xi=0.56$, as shown in Fig. \ref{Fig:delay_versus_user_density}. Firstly, simulation results and analytical bounds are highly consistent with each other. Secondly, the content retrieval delay increases with the growth of the network density, because both cellular and D2D transmission rates decrease in dense networks. For illustration, users spend about 23\% more time to retrieve contents when the network density increases from 80 to 400 per km$^2$ for $K=4$. More importantly, with high-rate cellular and D2D communications, the proposed caching policy with four SBSs reduces as much as 45\% content retrieval delay compared with the MPC policy.

As shown in Fig. \ref{Fig:delay_versus_backhaul_capacity}, we plot the average transmission delay in terms of the backhaul capacity. It can be observed that the performance gain acquired by the DCEC policy vanishes with the increase of backhaul capacity. When the constraint backhaul capacity acts as the bottleneck of the system performance, the proposed policy cooperatively cache more contents in edge networks to reduce the content retrieval delay. However, with unconstrained backhaul capacity, the performance gain degrades because users are able to fetch contents with low latency from remote servers.

Figure \ref{Fig:tradeoff} shows the impact of the SBS cluster size on the average content retrieval delay for $\lambda_{BS}=100,200,400$. As aforementioned, there exists a tradeoff between the transmission efficiency and the caching diversity. 
We observe that the average content retrieval delay decreases and then increases with the growth of SBS cluster size. When the SBS cluster size is too large, the benefit of cooperative edge caching vanishes. The reason is that a large SBS cluster size caches more popular contents while reduces the average SBS cluster transmission rate due to long physical distances to fetch contents. In addition, the tradeoff relationship between caching diversity and transmission performance indicates the optimal SBS cluster size exists. Specifically, the optimal cluster size is 7 for $\lambda_{BS}=100$, while when the network density increases to 400 per km$^2$, the optimal SBS cluster size decreases to 6. 

Furthermore, Fig. \ref{Fig:delay_versus_cluster_size} presents the optimal SBS cluster size with respect to the backhaul capacity. We observe that the optimal cluster size decreases with the increase of backhaul capacity, which means that a large SBS cluster size is preferred in the backhaul constrained scenario. The reason is that the increase of backhaul capacity alleviates the performance gain provided by the DCEC policy. For illustration, the optimal cluster size is 7 for $2$ Gbit/s, while when the backhaul capacity increases to $16$ Gbit/s, the optimal SBS cluster size decreases to 3.

\section{Conclusion and Future Work}\label{sec.Conclusion}
In this paper, a D2D-assisted cooperative edge caching has been proposed in mmWave dense networks. The closed-form expressions of backhaul offloading and content retrieval delay performance of the proposed policy have been obtained taking the practical directional antennas model and the network density into consideration. Both analytical and simulation results have revealed the average content retrieval delay increases with the network density. Besides, the tradeoff relationship between caching diversity and transmission efficiency has been investigated. Comparing with state-of-the-art MPC policy via extensive simulations, the proposed caching policy provides significant performance gains in both backhaul traffic offloading and content retrieval delay. For future works, as the D2D transmission degrades significantly with the increase of the network density, coordinated scheduling schemes will be investigated for concurrent D2D transmissions in mmWave dense networks. In addition, we will also study the optimal content placement to further minimize the content retrieval delay. 

\appendix
\subsection{Proof of Lemma \ref{Lemma: lemma4}}\label{AppA}
When a user is served by its D2D peer via mmWave D2D communications, the received signal power is given by
\begin{equation}
S_D=P_{U}(G_u^m)^2h_1C r_d^{-\alpha}.
\end{equation} 	


The received interference power which consists of the received signal from all the other D2D transmitters $\Phi_D$, is 
\begin{equation}
\begin{split}
I_D
&=\sum_{i\in \Phi_{D} } P_{U}G_u(\theta_{t,i}) G_u(\theta_{r,i}) h_iC r_i^{-\alpha}
\end{split}
\end{equation} 	
where $r_i$ is the distance between the user and $i$-th D2D interferers.  

Similarly, the average D2D transmission rate is lower bounded by
\begin{equation}\label{equ:D2D rate}
\begin{split}
\mathbb{E}[R_D]
&=\phi W\log_2\left(1+\frac{S_D}{I_D+\sigma^2}\right)\\
&\geq \frac{\phi W}{\ln 2}\left(\mathbb{E}[\ln S_D]-\ln \sum_{i\in \Phi_{D} }\mathbb{E}[I_D^i]\right)\\
&\geq \frac{\phi W}{\ln 2}\left(2\ln \frac{G_u^m}{\bar{G}_u}-\gamma-\alpha \mathbb{E}[\ln r_d]-\ln\sum_{i\in \Phi_{D} }\mathbb{E}[r_i^{-\alpha}]\right).
\end{split}
\end{equation}

Following the similar step in \eqref{equ: desiredSignal final version}, with PDF of $r_d$ in \eqref{equ:D2D distance distribution}, $\mathbb{E}[\ln r_d]$ is given by
\begin{equation}\label{equ:D2D_rd}
\begin{split}
 \mathbb{E}[\ln r_d]
&=\int_{0}^{r_d^{max}}\ln r_d \frac{2 r_d}{(r_d^{max})^2} dr_d\\
&=\ln r_d^{max}-\frac{1}{2}.
\end{split}
\end{equation}

Then, for the second term in \eqref{equ:D2D rate}, $\sum_{i\in \Phi_{D} }\mathbb{E}[r_i^{-\alpha}]$ is obtained in the following step. 
$r_i$ represents the inter-node distances of PPP, which follows a generalized beta distribution \cite{Haenggi2010distance}. 
The $-\alpha$th moment of $r_i$ is given by
\begin{eqnarray}
\mathbb{E}[r_i^{-\alpha}] = \left\{\begin{matrix}
\begin{split}
&\frac{R^{-\alpha}\Gamma(N_D+1)\Gamma(i-\frac{\alpha}{2})}{\Gamma(i)\Gamma(N_D+1-\frac{\alpha}{2})}, &\mbox{                 } \alpha <2\\ 
&\infty, &\mbox{ } \alpha \geq 2
\end{split}
\end{matrix}\right.
\end{eqnarray}
where $R=\sqrt{\frac{N_{D}}{\pi \lambda_{D}}}$ is the equivalent cell radius and $N_D$ is the number of D2D users. The summation of the $-\alpha$th moments of $r_i$ can be represented into different forms when $\alpha$ attains different values.

\begin{itemize}
	\item When $\alpha <2$,  the summation of moments of $r_i$ is given by
	\begin{equation}\label{equ:D2D sum moments a<2}
	\begin{split}
	\sum_{i\in \Phi_{D}}\mathbb{E}[r_i^{-\alpha}]
	&=\frac{R^{-\alpha}\Gamma(N_D+1)}{\Gamma(N_D+1-\frac{\alpha}{2})}\sum_{i=1}^{N_D}\frac{\Gamma(i-\frac{\alpha}{2})}{\Gamma(i)}\\
	&=\pi \lambda_{D}\frac{R^{2-\alpha}}{1-\frac{\alpha}{2}}
	\end{split}
	\end{equation}
	where the last the step due to the fact $\pi R^2\lambda_{D}=N_D$.
	
	\item When $\alpha >2$, the moment of $r_i$ becomes unbounded because the path loss model we adopt break down at short distances. Applying the guard radius $d_{0}$ around every receiver, i.e., the transmitters within $d_{0}$ are not allowed to transmit. The sum of interference within $d_{0}$ is $\pi \lambda_{D}{d_0^{2-\alpha}}/({1-\frac{\alpha}{2}})$ according to \eqref{equ:D2D sum moments a<2}. Excluding the interference within the guard radius, we have
	\begin{equation}\label{equ:D2D sum moments a>2}
	\begin{split}
	\sum_{i\in \Phi_{D} }\mathbb{E}[r_i^{-\alpha}]
	&=\pi \lambda_{D}\frac{R^{2-\alpha}-d_0^{2-\alpha}}{1-\frac{\alpha}{2}}.
	\end{split}
	\end{equation} 
	
	\item When $\alpha=2$, taking the limit of \eqref{equ:D2D sum moments a>2}, we obtain
	\begin{equation}\label{equ:D2D sum moments a=2}
	\begin{split}
	\sum_{i\in \Phi_{D} }\mathbb{E}[r_i^{-\alpha}]
	&=2\pi \lambda_{D} \ln \left(\frac{R}{d_0}\right).
	\end{split}
	\end{equation}
\end{itemize}

Substituting \eqref{equ:D2D_rd}, \eqref{equ:D2D sum moments a<2}, \eqref{equ:D2D sum moments a>2} and \eqref{equ:D2D sum moments a=2} into \eqref{equ:D2D rate}, {Lemma~\ref{Lemma: lemma4}} is proved.

\bibliographystyle{IEEEtran}
\bibliography{security}

\begin{thebibliography}{10}
\providecommand{\url}[1]{#1}
\csname url@samestyle\endcsname
\providecommand{\newblock}{\relax}
\providecommand{\bibinfo}[2]{#2}
\providecommand{\BIBentrySTDinterwordspacing}{\spaceskip=0pt\relax}
\providecommand{\BIBentryALTinterwordstretchfactor}{4}
\providecommand{\BIBentryALTinterwordspacing}{\spaceskip=\fontdimen2\font plus
\BIBentryALTinterwordstretchfactor\fontdimen3\font minus
  \fontdimen4\font\relax}
\providecommand{\BIBforeignlanguage}[2]{{%
\expandafter\ifx\csname l@#1\endcsname\relax
\typeout{** WARNING: IEEEtran.bst: No hyphenation pattern has been}%
\typeout{** loaded for the language `#1'. Using the pattern for}%
\typeout{** the default language instead.}%
\else
\language=\csname l@#1\endcsname
\fi
#2}}
\providecommand{\BIBdecl}{\relax}
\BIBdecl

\bibitem{xiao2017Jsac}
M.~Xiao, S.~Mumtaz, Y.~Huang, L.~Dai, Y.~Li, M.~Matthaiou, G.~K. Karagiannidis,
  E.~Bj{\"o}rnson, K.~Yang, and I.~Chih-Lin, ``Millimeter wave communications
  for future mobile networks,'' \emph{IEEE J. Sel. Areas Commun.}, vol.~35,
  no.~9, pp. 1909--1935, 2017.

\bibitem{cheng2018big}
N.~Cheng, F.~Lyu, J.~Chen, W.~Xu, H.~Zhou, S.~Zhang, and X.~Shen, ``Big data
  driven vehicular networks,'' \emph{IEEE Network}, no.~99, pp. 1--8,
  DOI.10.1109/MNET.2018.1700460, 2018.

\bibitem{rappaport2013it_will_work}
T.~S. Rappaport, S.~Sun, R.~Mayzus, H.~Zhao, Y.~Azar, K.~Wang, G.~N. Wong,
  J.~K. Schulz, M.~Samimi, and F.~Gutierrez, ``Millimeter wave mobile
  communications for {5G} cellular: {It} will work!'' \emph{IEEE Access},
  vol.~1, pp. 335--349, 2013.

\bibitem{cheng2017performance}
N.~Cheng, H.~Zhou, L.~Lei, N.~Zhang, Y.~Zhou, X.~Shen, and F.~Bai,
  ``Performance analysis of vehicular device-to-device underlay
  communication,'' \emph{IEEE Trans. Veh. Technol.}, vol.~66, no.~6, pp.
  5409--5421, 2017.

\bibitem{bastug2014living}
E.~Bastug, M.~Bennis, and M.~Debbah, ``Living on the edge: {The} role of
  proactive caching in {5G} wireless networks,'' \emph{IEEE Commun. Mag.},
  vol.~52, no.~8, pp. 82--89, 2014.

\bibitem{zhong2017towards}
Y.~Zhong, M.~Haenggi, F.~Zheng, W.~Zhang, T.~Q. Quek, and W.~Nie, ``Towards a
  tractable delay analysis in ultra-dense networks,'' \emph{IEEE Commun. Mag.},
  vol.~55, no.~12, pp. 103--109, 2017.

\bibitem{ye2017distributed}
Q.~Ye, W.~Zhuang, L.~Li, and P.~Vigneron, ``Traffic-load-adaptive medium access
  control for fully connected mobile ad hoc networks,'' \emph{IEEE Trans. Veh.
  Technol.}, vol.~65, no.~11, pp. 9358--9371, 2016.

\bibitem{edgeCachingWhitePaper}
``Mobile-edge computing - introductory technical white paper,'' \emph{European
  Telecommunications Standards Institute, Tech.}, Rep., Sep. 2014, accessed
  Mar. 27, 2017.[Online]. Available: https: //portal.etsi.org/.

\bibitem{SS-MAC}
F.~Lyu, H.~Zhu, H.~Zhou, W.~Xu, N.~Zhang, M.~Li, and X.~Shen, ``{SS-MAC}: {A}
  novel time slot-sharing {MAC} for safety messages broadcasting in {VANETs},''
  \emph{IEEE Trans. Veh. Technol.}, vol.~67, no.~4, pp. 3586--3597, 2018.

\bibitem{rodrigues2017hybrid}
T.~G. Rodrigues, K.~Suto, H.~Nishiyama, and N.~Kato, ``Hybrid method for
  minimizing service delay in edge cloud computing through {VM} migration and
  transmission power control,'' \emph{IEEE Trans. Comput.}, vol.~66, no.~5, pp.
  810--819, 2017.

\bibitem{rodrigues2018cloudlets}
T.~G. Rodrigues, K.~Suto, H.~Nishiyama, N.~Kato, and K.~Temma, ``Cloudlets
  activation scheme for scalable mobile edge computing with transmission power
  control and virtual machine migration,'' \emph{IEEE Trans. Comput.}, vol.~67,
  no.~9, pp. 1287--1300, 2018.

\bibitem{zhou2017resource}
Y.~Zhou, F.~R. Yu, J.~Chen, and Y.~Kuo, ``Resource allocation for
  information-centric virtualized heterogeneous networks with in-network
  caching and mobile edge computing,'' \emph{IEEE Trans. Veh. Technol.},
  vol.~66, no.~12, pp. 11\,339--11\,351, 2017.

\bibitem{liu2016outage}
J.~Liu, H.~Nishiyama, N.~Kato, and J.~Guo, ``On the outage probability of
  device-to-device-communication-enabled multichannel cellular networks: {An
  RSS}-threshold-based perspective.'' \emph{IEEE J. Sel. Areas Commun.},
  vol.~34, no.~1, pp. 163--175, 2016.

\bibitem{song2018stable}
W.~Song, Y.~Zhao, and W.~Zhuang, ``Stable device pairing for collaborative data
  dissemination with device-to-device communications,'' \emph{IEEE Internet of
  Things J.}, vol.~5, no.~2, pp. 1251--1264, 2018.

\bibitem{zhou2017energy}
Z.~Zhou, K.~Ota, M.~Dong, and C.~Xu, ``Energy-efficient matching for resource
  allocation in {D2D} enabled cellular networks,'' \emph{IEEE Trans. Veh.
  Technol.}, vol.~66, no.~6, pp. 5256--5268, 2017.

\bibitem{ji2015throughput}
M.~Ji, G.~Caire, and A.~F. Molisch, ``The throughput-outage tradeoff of
  wireless one-hop caching networks,'' \emph{IEEE Trans. Inf. Theory}, vol.~61,
  no.~12, pp. 6833--6859, 2015.

\bibitem{wang2017mobility}
R.~Wang, J.~Zhang, S.~Song, and K.~B. Letaief, ``Mobility-aware caching in
  {D2D} networks,'' \emph{IEEE Trans. Wireless Commun.}, vol.~16, no.~8, pp.
  5001--5015, 2017.

\bibitem{zhao2018caching}
N.~Zhao, X.~Liu, Y.~Chen, S.~Zhang, Z.~Li, B.~Chen, and M.~S. Alouini,
  ``Caching {D2D} connections in small-cell networks,'' \emph{IEEE Trans. Veh.
  Technol.}, DOI: 10.1109/TVT.2018.2877645, 2018.

\bibitem{chen2017cooperative}
Z.~Chen, J.~Lee, T.~Q. Quek, and M.~Kountouris, ``Cooperative caching and
  transmission design in cluster-centric small cell networks,'' \emph{IEEE
  Trans. Wireless Commun.}, vol.~16, no.~5, pp. 3401--3415, 2017.

\bibitem{zhang2017cooperative}
S.~Zhang, P.~He, K.~Suto, P.~Yang, L.~Zhao, and X.~Shen, ``Cooperative edge
  caching in user-centric clustered mobile networks,'' \emph{IEEE Trans. Mobile
  Comput.}, vol.~17, no.~8, pp. 1791--1805, 2018.

\bibitem{zhang2017cost}
S.~Zhang, N.~Zhang, P.~Yang, and X.~Shen, ``Cost-effective cache deployment in
  mobile heterogeneous networks,'' \emph{IEEE Trans. Veh. Technol.}, vol.~66,
  no.~12, pp. 11\,264--11\,276, 2017.

\bibitem{xu2018saving}
J.~Xu, K.~Ota, and M.~Dong, ``Saving energy on the edge: {In}-memory caching
  for multi-tier heterogeneous networks,'' \emph{IEEE Commun. Mag.}, vol.~56,
  no.~5, pp. 102--107, 2018.

\bibitem{zhao2018collaborative}
X.~Zhao, P.~Yuan, H.~Li, and S.~Tang, ``Collaborative edge caching in
  context-aware device-to-device networks,'' \emph{IEEE Trans. Veh. Technol.},
  vol.~67, no.~10, pp. 9583--9596, 2018.

\bibitem{yang2018content}
P.~Yang, N.~Zhang, S.~Zhang, L.~Yu, J.~Zhang, and X.~Shen, ``Content popularity
  prediction towards location-aware mobile edge caching,'' \emph{IEEE Trans.
  Multimedia}, DOI 10.1109/TMM.2018.2870521, 2018.

\bibitem{song2016packet}
W.~Song and W.~Zhuang, ``Packet assignment under resource constraints with
  {D2D} communications,'' \emph{IEEE Network}, vol.~30, no.~5, pp. 54--60,
  2016.

\bibitem{semiari2018caching}
O.~Semiari, W.~Saad, M.~Bennis, and B.~Maham, ``Caching meets millimeter wave
  communications for enhanced mobility management in {5G} networks,''
  \emph{IEEE Trans. Wireless Commun.}, vol.~17, no.~2, pp. 779--793, 2018.

\bibitem{ji2016wireless}
M.~Ji, G.~Caire, and A.~F. Molisch, ``Wireless device-to-device caching
  networks: {Basic} principles and system performance,'' \emph{IEEE J. Sel.
  Areas Commun.}, vol.~34, no.~1, pp. 176--189, 2016.

\bibitem{giatsoglou2017d2d-aware}
N.~Giatsoglou, K.~Ntontin, E.~Kartsakli, A.~Antonopoulos, and C.~Verikoukis,
  ``{D2D}-aware device caching in mmwave-cellular networks,'' \emph{IEEE J.
  Sel. Areas Commun.}, vol.~35, no.~9, pp. 2025--2037, 2017.

\bibitem{zhong2017heterogeneous}
Y.~Zhong, T.~Q. Quek, and X.~Ge, ``Heterogeneous cellular networks with
  spatio-temporal traffic: {Delay} analysis and scheduling,'' \emph{IEEE J.
  Sel. Areas Commun.}, vol.~35, no.~6, pp. 1373--1386, 2017.

\bibitem{singh2011interference}
S.~Singh, R.~Mudumbai, and U.~Madhow, ``Interference analysis for highly
  directional 60-{GHz} mesh networks: {The} case for rethinking medium access
  control,'' \emph{IEEE/ACM Trans. Netw.}, vol.~19, no.~5, pp. 1513--1527,
  2011.

\bibitem{maccartney2017rural}
G.~R. MacCartney and T.~S. Rappaport, ``Rural macrocell path loss models for
  millimeter wave wireless communications,'' \emph{IEEE J. Sel. Areas Commun.},
  vol.~35, no.~7, pp. 1663--1677, 2017.

\bibitem{wu2017Wiopt}
W.~Wu, Q.~Shen, K.~Aldubaikhy, N.~Cheng, N.~Zhang, and X.~Shen, ``Enhance the
  edge with beamforming: {Performance} analysis of beamforming-enabled
  {WLAN},'' in \emph{Proc. IEEE WiOpt}, 2018.

\bibitem{bai2015coverage}
T.~Bai and R.~W. Heath, ``Coverage and rate analysis for millimeter-wave
  cellular networks,'' \emph{IEEE Trans. Wireless Commun.}, vol.~14, no.~2, pp.
  1100--1114, 2015.

\bibitem{yu2013downlink}
S.~M. Yu and S.~L. Kim, ``Downlink capacity and base station density in
  cellular networks,'' in \emph{Proc. IEEE WiOpt}, 2013, pp. 119--124.

\bibitem{haenggi2009stochastic}
M.~Haenggi, J.~G. Andrews, F.~Baccelli, O.~Dousse, and M.~Franceschetti,
  ``Stochastic geometry and random graphs for the analysis and design of
  wireless networks,'' \emph{IEEE J. Sel. Areas Commun.}, vol.~27, no.~7, pp.
  1029--1046, 2009.

\bibitem{Haenggi2010distance}
S.~Srinivasa and M.~Haenggi, ``Distance distributions in finite uniformly
  random networks: {Theory} and applications,'' \emph{IEEE Trans. Veh.
  Technol.}, vol.~59, no.~2, pp. 940--949, 2010.

\bibitem{wu2017performance}
W.~Wu, Q.~Shen, M.~Wang, and X.~Shen, ``Performance analysis of {IEEE} 802.11.
  ad downlink hybrid beamforming,'' in \emph{Proc. IEEE ICC}, 2017.

\end{thebibliography}

\end{document}